\documentclass[twocolumn]{emulateapj}
\usepackage{amsmath}
\usepackage{amssymb}
\usepackage{amstext}
\usepackage{epsfig}
\usepackage{graphicx}

\begin{document}
\shorttitle{X1908+075}
\shortauthors{LEVINE et al.}
\title{X1908+075: A Pulsar Orbiting in the Stellar Wind of a Massive Companion}

\author{A.\ M.\ Levine\altaffilmark{1}, S. \ Rappaport\altaffilmark{1},
R.\ Remillard\altaffilmark{1}, \& A.\ Savcheva\altaffilmark{1}}

\altaffiltext{1}{Department of Physics and Center for Space Research,
MIT 37-575, 77 Massachusetts Ave, Cambridge, MA 02139; 
{\tt aml@space.mit.edu}}

\begin{abstract}

We have observed the persistent but optically unidentified X-ray
source X1908+075 with the PCA and HEXTE instruments on {\it RXTE}.
The binary nature of this source was established by \citet{WRB} who
found a 4.4-day orbital period in results from the {\it RXTE} ASM.  We
report the discovery of 605 s pulsations in the X-ray flux.  The
Doppler delay curve is measured and provides a mass function of 6.1
$M_\sun$ which is a lower limit to the mass of the binary companion of
the neutron star.  The degree of attenuation of the low-energy end of
the spectrum is found to be a strong function of orbital phase.  A
simple model of absorption in a stellar wind from the companion star
fits the orbital phase dependence reasonably well and limits the
orbital inclination angle to the range $38\arcdeg - 72\arcdeg$.  These
measured parameters lead to an orbital separation of $\sim60 - 80$
lt-s, a mass for the companion star in the range 9-31 $M_\sun$, and an
upper limit to the size of the companion of $\sim 22~R_\sun$.  From
our analysis we also infer a wind mass loss rate from the companion
star of $\gtrsim 1.3 \times 10^{-6}~M_\sun$ yr$^{-1}$ and, when the
properties of the companion star and the effects of photoionization
are considered, likely $\gtrsim 4 \times 10^{-6}~M_\sun$ yr$^{-1}$.
Such a high rate is inconsistent with the allowed masses and radii
that we find for a main sequence or modestly evolved star unless the
mass loss rate is enhanced in the binary system relative to that of an
isolated star.  We discuss the possibility that the companion might be
a Wolf-Rayet star that could evolve to become a black hole in $10^4$
to $10^5$ yr.  If so, this would be the first identified progenitor of
a neutron star--black hole binary.

\end{abstract}

 
\section{INTRODUCTION}

X1908+075 is an optically unidentified, highly absorbed, and
relatively faint X-ray source that appeared in surveys carried out
with instruments on the {\em Uhuru}, {\em OSO 7}, {\em Ariel 5}, {\em
HEAO-1}, and {\em EXOSAT} satellites.  The early detections and
position determinations are summarized by \citet[hereafter WRB]{WRB}.
They conclude that the position of X1908+075 is likely to be within
the overlapping region of the error box of a source detected in an
{\em Einstein} IPC image and one of an array of {\em HEAO 1} A-3
position ``diamonds'', and is thus known with an accuracy of $\sim
1'$. Inspection of the POSS plates within the source error box reveals
no optical counterpart down to magnitude 20.  This is consistent with
the heavy optical extinction implied by the interstellar hydrogen
column density of $\sim 4 \times 10^{22}$ atoms cm$^{-2}$ measured on
the basis of the low-energy absorption in the X-ray spectrum
\citep[see][]{ss85}.

The intensity of X1908+075 has been monitored for the past eight years
using the All-Sky Monitor (ASM) aboard the {\it Rossi X-Ray Timing
Explorer} ({\it RXTE}), and has typically been in the range 2-12 mCrab
in the 2-12 keV energy band.  WRB analyzed data from the first three
years of operation of the ASM, and thereby discovered a 4.4 day
periodicity in the X-ray intensity.  The periodic component of the
intensity variations is clearly energy dependent in both strength and
detailed form.  At most orbital phases, the variation is roughly
sinusoidal while a relatively sharp dip forms the minimum in the 5-12
keV band.  These characteristics suggest that the modulation is
produced by a varying amount of absorption along the line of sight as
the source moves through the stellar wind of a massive companion
star. The hard X-ray spectrum led WRB to also suggest that this source
could be an X-ray pulsar.

The possibility that X1908+075 might be an X-ray pulsar led us to
carry out a set of pointed observations with the PCA and HEXTE
instruments on {\it RXTE} in late 2000 and early 2001.  The data
revealed the presence of strong X-ray pulsations at a period of 605
seconds.  We were also able to detect Doppler delays in the pulse
arrival times.  However, because the number of independent
high-quality pulse arrival times obtained from this data set was
small, it proved difficult to unambiguously disentangle orbital
effects from intrinsic changes in the pulse period.  The latter could,
in principle, be quite large for a neutron star rotating with a period
as long as 600~s \citep[see, e.g.,][]{bild97,del01}.  Therefore, we
obtained additional observations of the source with {\it RXTE} during
late 2002 and early 2003.

In this paper we report the results of our analysis of the {\it RXTE}
observations of X1908+075.  The pointed observations are described in
\S 2.  A pulse timing analysis is described in \S 3.  We present an
accurate pulse period and a determination of the spin-down of the
neutron star due to accretion and magnetic torques.  The orbital
Doppler delay curve is measured, thereby confirming the 4.4-day period
found in the ASM X-ray light curve.  The resultant mass function is
$6.1~M_\sun$, indicating that the pulsar does indeed orbit a massive
companion star.  The results of an orbital-phase-dependent spectral
analysis are presented in \S 4.  We detect a very pronounced
modulation of the low energy attenuation as a function of orbital
phase.  In \S 5 we model this modulation by absorption in a
spherically symmetric stellar wind, whereby we obtain constraints on
the orbital inclination, on the properties of the companion star, and
on the stellar wind.  We discuss the implications of our results in \S
6, including the possibility that this system may be the progenitor of
a neutron star--black hole binary.

\section{OBSERVATIONS}

Pointed observations of X1908+075 were made with the Proportional
Counter Array (PCA) and High-Energy X-ray Timing Experiment (HEXTE) on
{\it RXTE}.  The PCA consists of 5 Proportional Counter Units (PCUs)
that are sensitive to X-ray photons in the range 2.5--60~keV.  Each
PCU has a collecting area of $\sim 1400\ {\rm cm^2}$ and a collimator
to limit the field of view to $1\arcdeg$ in radius \citep{xte96}.
Some of the PCUs are operated with reduced duty cycle in order to
avoid problems associated with constant use.  Only PCUs 0 and 2 were
used in every observation.  Data from the PCA observations were
telemetered to the ground in the ``GoodXenon'' mode, which includes
information on each good event with 1 $\mu$s time resolution and the
instrument's full energy resolution (255 channels).  The HEXTE
comprises two clusters, each of which includes 4 NaI scintillation
detectors sensitive to photons in the range 15--250~keV that provide a
total collecting area of $800\ {\rm cm^2}$ per cluster.  The detectors
in each cluster view a common $1\arcdeg$ radius field \citep{Roth98}.

Between 2000 November 19 and 2001 February 22, we obtained 15 short
observations with a typical exposure time in each of $\sim 2000$ s.  A
total exposure of 35 ks was obtained in these observations, which we
refer to collectively as ``epoch 1''.  Over all of the 15 observations
there were, on average, 3.0 PCUs in operation. Eight of the
observations were made during one orbital cycle of X1908+075 ($\sim 4$
days), while the remaining 7 were performed nearly 3 months later.
Fourier transforms of these data clearly exhibited the presence of
X-ray pulsations with a period of 605 s.

\begin{figure*}[t]
\begin{center}
\includegraphics[width=5.0in,angle=90]{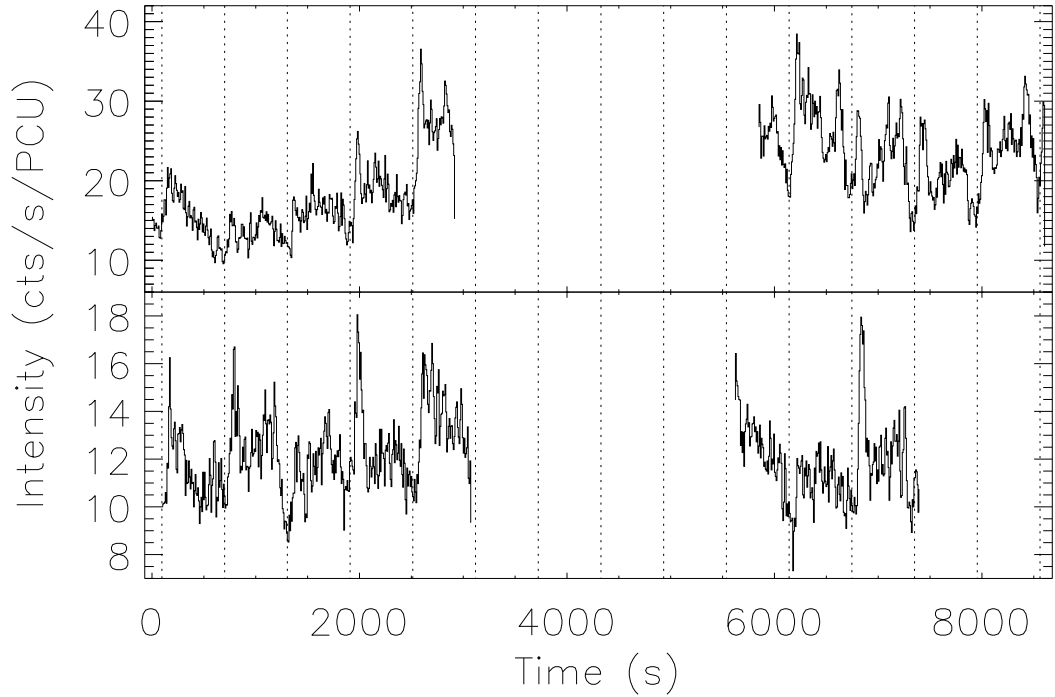}
\caption{PCA counting rate data for the 3.7--17 keV energy band in
8-s time bins for two representative observations of X1908+075, i.e.,
the observations of 2002 December 23 (bottom; Obs. ID 70083-01-01-00)
and of 2003 February 8 (top; Obs. ID 70083-02-19-00).  The vertical
lines are drawn every 605 seconds and are roughly aligned with the
principal minimum in the pulse profile.  The gap in the middle of each
observation is due to Earth occultation of the source. The non-source
background rate is $\sim 7.5$ cts s$^{-1}$ PCU$^{-1}$.
\label{fig:cnts}}
\end{center}
\end{figure*}

Additional observations with the PCA and HEXTE were carried out on 39
occasions from 2002 December 23 through 2003 January 5 and on 32
occasions from 2003 January 30 through 2003 February 8 (``epoch 2'';
196 ks total exposure).  On average, 3.5 PCUs were in operation.  Each
of these 71 observations in epoch 2 typically yielded $\sim 3000$ s of
net exposure most often in the form of two contiguous time intervals
separated by a $\sim 2000$ s gap due to Earth occultation of the
target.  Raw counting rate data from two of these 71 pointed
observations are shown in Fig. \ref{fig:cnts} to illustrate the
typical appearance of the X-ray pulsations and accompanying source
variability.  The X-ray intensity, in addition to exhibiting obvious
pulsations, is also quite variable on timescales comparable to the
pulse period and longer.

For pulsar timing purposes we used events with energies in the range
3.7--17 keV (channels 3--40) from both left and right sections of
layer 1 of all operating PCUs.  Event times were reduced to the Solar
System barycenter.  The average 3.7--17 keV background count rate in
individual observations done in 2003-2003 ranges from 7.3 to 7.8 cts
s$^{-1}$ PCU$^{-1}$.  Background was not subtracted in our timing
analyses.

To determine an appropriate model for spectral analyses, we used data
from both the PCA and HEXTE.  In the PCA analyses we utilize 2.9--25
keV pulse height range data from all layers of PCU 2 only because that
PCU was operational for all of the observations and because its
calibration is superior, as judged by the analysis of contemporaneous
PCA observations of the Crab Nebula. Spectral extractions and
background subtractions for both the PCA and HEXTE were performed
using the ``FTOOLS'' package, and spectral models were applied using
``XSPEC''.  Both of these software packages were provided by
NASA/HEASARC.


\section{PULSE TIMING AND ORBITAL ANALYSIS}

A first estimate of the mean pulse period was obtained by binning the
data from all of the epoch 2 (2002-2003) observations in 16-s bins and
carrying out an FFT analysis of the resultant data train ($2^{18}$
points).  The FFT revealed a highly significant signal at a period of
604.76 s, as well as 5 very prominent higher harmonics of the
fundamental.  Also evident in the Fourier transform were a number of
sidebands of the pulsations that were spaced in frequency by multiples
of 1/4.4 days$^{-1}$.  These sidebands were identified as due to the
orbital motion of the neutron star about its companion.  From the
fundamental and the prominent harmonics and sidebands, we derived an
average pulse period during these observations of 604.689 s.

\begin{figure*}[t]
\begin{center}
\includegraphics[width=5.0in,angle=90]{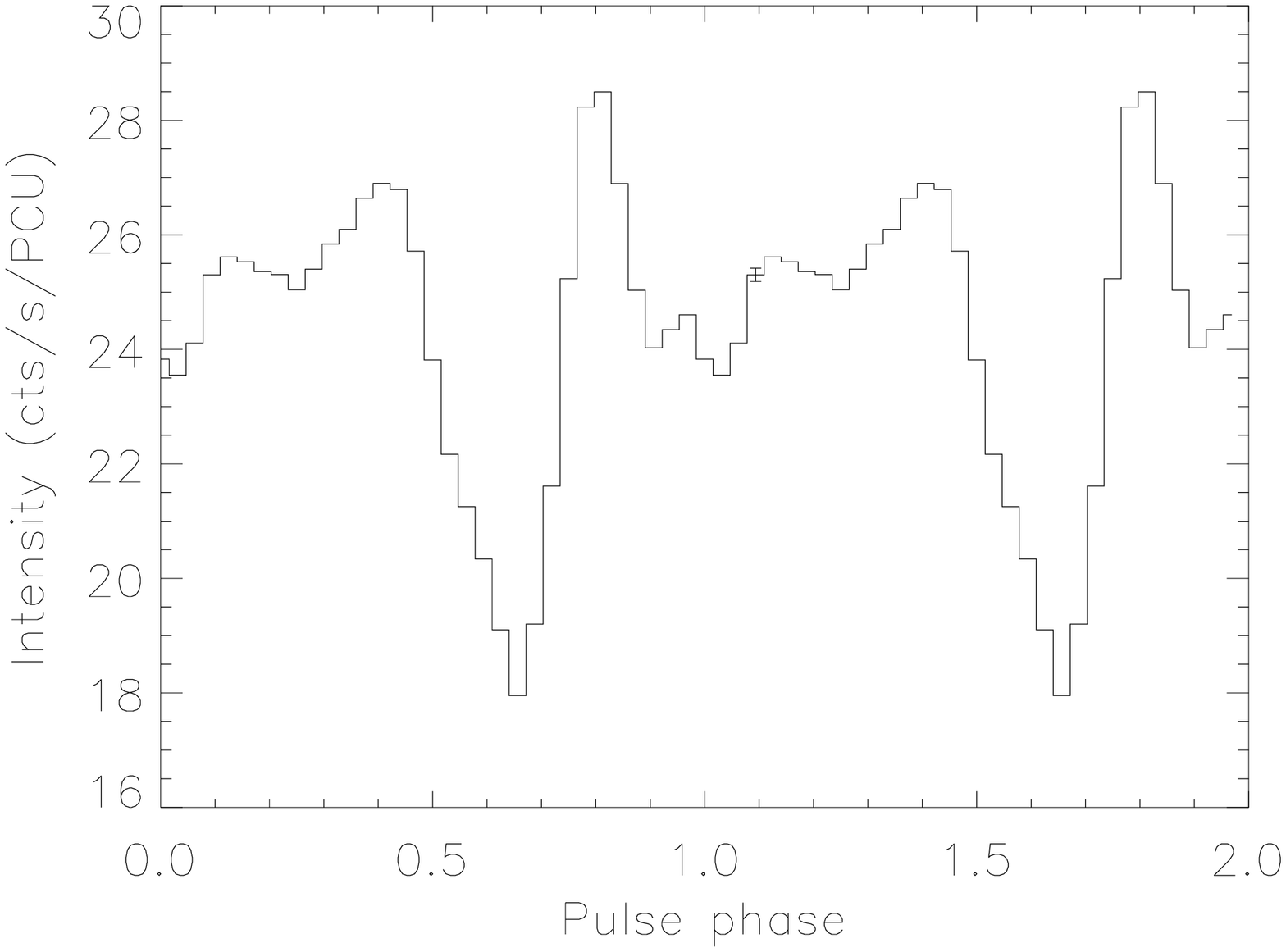}
\caption{Pulse profile template used for computing the cross
correlation function. The profile is shown for two cycles.  The error
bar shows the approximate uncertainty ($\pm 1\ \sigma$) from
variations of the source intensity at a given pulse phase within a
typical observation.  Fluctuations due to counting statistics are
negligible.  The non-source background rate is $\sim 7.5$ cts s$^{-1}$
PCU$^{-1}$.
\label{fig:pulse}}
\end{center}
\end{figure*}

We proceeded to fold the data from each of the 71 pointed observations
from 2002-2003 modulo this value of the average pulse period.  To form
a ``pulse template'', the individual pulse profiles were manually
aligned to correct for small phase drifts and then averaged (see
Fig. \ref{fig:pulse}).


\begin{figure*}[t]
\begin{center}
\includegraphics[width=5.0in,angle=90]{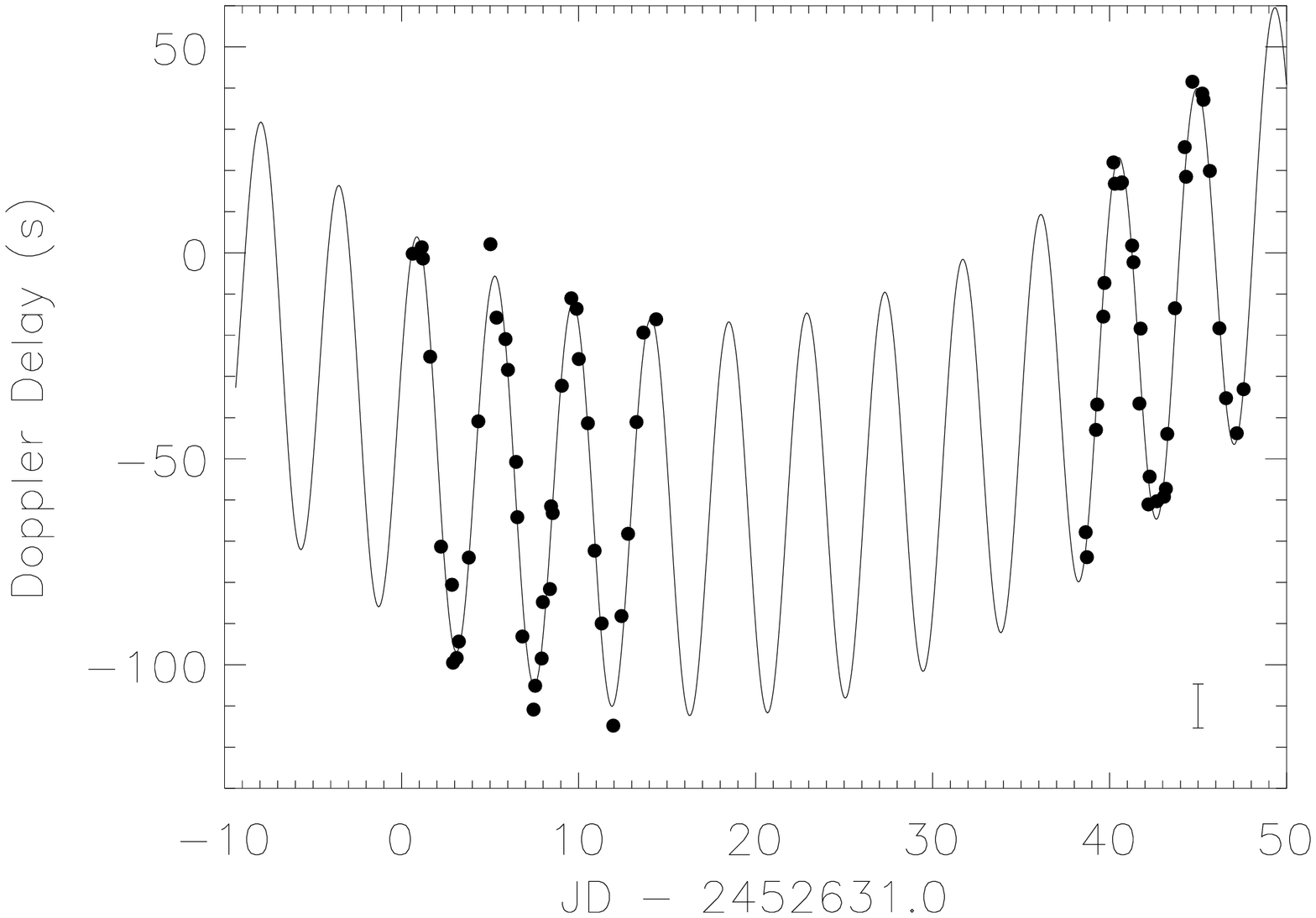}
\caption{Pulse arrival time delays for 69 pointed observations made
during the interval 2002 December 23 through 2003 February 8.  The
delays are computed relative to a reference pulse ephemeris that has a
constant 604.689 s period at the Solar System barycenter.  The curve
shows the arrival time delays predicted by the best-fit model pulsar
with a constant pulse period derivative moving in a circular orbit
(see Table \ref{timeparam}).  A $\pm 1~\sigma$ error bar, estimated
from the rms deviation of the measurements relative to the curve, is
plotted in the lower right corner.
\label{fig:delay}}
\end{center}
\end{figure*}

We then computed the cross correlation function (CCF) between the
template and the folded profile from each of the individual
observations.  The phase difference for which the CCF reaches a
maximum value is the best estimate of the pulsar phase relative to the
folding ephemeris.  Pulse arrival time delays are then simply obtained
as the products of the phase differences and the folding period.
Pulse arrival times were then computed by adding each time delay to
the time of phase zero of a pulse near the middle of the observation
interval.  This analysis yielded a total of 69 pulse arrival times
after we eliminated the results from 2 of the 71 observations for
which the CCFs were possibly ambiguous.  The pulse arrival time delays
are listed in Table \ref{tab:pulse} and plotted in Figure
\ref{fig:delay}.  The Doppler delays due to the orbital motion are
quite apparent in Figure \ref{fig:delay}, as is an overall quadratic
behavior due to the slowdown in the rotation rate of the neutron star.

We fit the pulse arrival times with a 7-parameter model of the orbit
and pulse period behavior.  The arrival time of the $n$th pulse is given
by
\begin{eqnarray}
t_n & = & t_0+nP+\frac{1}{2}n^2P\dot P + \frac{1}{c} a_x \sin i \cos[\Omega (t_n-\tau_{90})]
 \nonumber \\
 & & - \frac{e}{2c} a_x \sin i \sin[2\Omega (t_n-\tau_{90})-\omega_p]
\end{eqnarray}
where $P$ is the pulse period at time $t_0$, $\dot P$ is the pulse
period derivative, $a_x \sin i$ is the projected semimajor axis of the
orbit of the neutron star, $\tau_{90}$ is a reference time, which for
a circular orbit corresponds to superior conjunction, $\Omega$ is the
orbital frequency which we fix at a value of $2\pi/4.400$ radians
day$^{-1}$, and $\omega_p$ is the longitude of periastron.  The last
term represents the first order term in a Taylor series expansion in
the eccentricity $e$ and is a reasonable approximation for a mildly
eccentric orbit.  The curve representing the best-fit orbit and pulse
period behavior is shown superposed on the arrival time delays in
Fig. \ref{fig:delay}.  The rms scatter of the arrival time delays
about this best fit curve was found to be $\sim 6$ s and was used as
an empirical estimate of the uncertainty in the individual arrival
times.  This was the basis on which we derived formal confidence
limits on the individual fitted parameters.

\begin{deluxetable*}{cccccc}
\tablecaption{X1908+075 Pulse Arrival Time Delays}
\tablehead{
\colhead{Observation} & \colhead{Time of Obs.\tablenotemark{a}} & \colhead{Exposure} &
\colhead{Count Rate\tablenotemark{b}}
 & \colhead{Orb. Phase\tablenotemark{c}} & \colhead{Pulse Time\tablenotemark{d}} \\
\colhead{ID} & \colhead{(MJD)} & \colhead{(s)} & 
\colhead{(cts s$^{-1}$)} & \colhead{(cycles)} & \colhead{Delay (s)}
}

\startdata
70083-01-01-00  &  52631.1289  &  3344  &   7.79 &  0.942 & $-0.2$ \\
70083-01-02-00  &  52631.6400  &  1488  &  12.30 &  0.058 & $1.4$ \\
70083-01-02-01  &  52631.7077  &  1440  &  21.77 &  0.074 & $-1.4$ \\
70083-01-03-00  &  52632.1150  &  3424  &  16.24 &  0.166 & $-25.2$ \\
70083-01-04-00  &  52632.7296  &  1200  &   8.38 &  0.306 & $-71.3$ \\
70083-01-05-00  &  52633.3469  &  2448  &  15.19 &  0.446 & $-80.6$ \\
70083-01-05-01  &  52633.4084  &  1312  &  22.48 &  0.460 & $-99.5$ \\
70083-01-06-01  &  52633.6143  &  1856  &  12.29 &  0.507 & $-98.4$ \\
70083-01-06-00  &  52633.7382  &  3216  &  14.14 &  0.535 & $-94.4$ \\
70083-01-07-00  &  52634.2994  &  2816  &  10.10 &  0.663 & $-74.0$ \\
70083-01-08-00  &  52634.8365  &  1200  &   9.68 &  0.785 & $-40.9$ \\
70083-01-09-01  &  52635.5223  &  2016  &   5.12 &  0.941 & $2.1$ \\
70083-01-10-00  &  52635.8639  &  3312  &   8.31 &  0.018 & $-15.7$ \\
70083-01-11-01  &  52636.3726  &  2080  &   7.18 &  0.134 & $-20.9$ \\
70083-01-11-00  &  52636.5098  &  2032  &   4.96 &  0.165 & $-28.4$ \\
70083-01-12-01  &  52636.9702  &  1888  &   7.55 &  0.270 & $-50.7$ \\
70083-01-12-00  &  52637.0396  &  1520  &  10.89 &  0.285 & $-64.2$ \\
70083-01-13-00  &  52637.3282  &  2784  &  12.79 &  0.351 & $-93.1$ \\
70083-01-14-00  &  52637.9570  &  1776  &  13.97 &  0.494 & $-110.9$ \\
70083-01-14-01  &  52638.0480  &  1408  &  29.83 &  0.515 & $-105.1$ \\
70083-01-15-00  &  52638.4150  &  2240  &  14.51 &  0.598 & $-98.5$ \\
70083-01-15-01  &  52638.4827  &  2384  &  17.15 &  0.613 & $-84.8$ \\
70083-01-16-01  &  52638.8809  &  1008  &  16.73 &  0.704 & $-81.6$ \\
70083-01-16-00  &  52638.9476  &  1008  &  23.72 &  0.719 & $-61.5$ \\
70083-01-16-02  &  52639.0372  &  1168  &  19.72 &  0.739 & $-63.2$ \\
70083-01-17-00  &  52639.5570  &  2864  &  12.72 &  0.857 & $-32.3$ \\
70083-01-18-00  &  52640.0893  &  3232  &  11.46 &  0.978 & $-11.0$ \\
70083-01-19-00  &  52640.3903  &  2672  &  11.24 &  0.047 & $-13.6$ \\
70083-01-19-01  &  52640.5146  &  1904  &  17.32 &  0.075 & $-25.8$ \\
70083-01-20-00  &  52641.0247  &  2960  &  14.42 &  0.191 & $-41.4$ \\
70083-01-21-00  &  52641.4070  &  2720  &  21.03 &  0.278 & $-72.3$ \\
70083-01-22-00  &  52641.8042  &  2192  &  12.16 &  0.368 & $-90.0$ \\
70083-01-23-00  &  52642.4619  &  2816  &  16.83 &  0.518 & $-114.8$ \\
70083-01-24-00  &  52642.9233  &  1632  &  12.85 &  0.622 & $-88.2$ \\
70083-01-25-00  &  52643.2959  &  1088  &  15.11 &  0.707 & $-68.2$ \\
70083-01-26-00  &  52643.7705  &  2608  &  12.80 &  0.815 & $-41.1$ \\
70083-01-27-00  &  52644.1608  &  3408  &   8.86 &  0.904 & $-19.4$ \\
70083-01-28-00  &  52644.8823  &  2624  &   9.46 &  0.068 & $-16.1$ \\
70083-02-01-00  &  52669.1497  &  3536  &  13.09 &  0.582 & $-67.8$ \\
70083-02-01-01  &  52669.2184  &  1808  &  17.62 &  0.598 & $-73.9$ \\
70083-02-02-00  &  52669.7313  &  1648  &  17.05 &  0.714 & $-43.0$ \\
70083-02-02-01  &  52669.7997  &  2080  &  10.21 &  0.730 & $-36.8$ \\
70083-02-03-00  &  52670.1434  &  2112  &  19.80 &  0.808 & $-15.5$ \\
70083-02-03-01  &  52670.2066  &  1904  &  13.94 &  0.822 & $-7.3$ \\
70083-02-04-00  &  52670.7184  &  1600  &  11.88 &  0.938 & $22.0$ \\
70083-02-04-01  &  52670.7955  &  1952  &  13.91 &  0.956 & $16.8$ \\
70083-02-05-00  &  52671.1230  &  2640  &   8.26 &  0.030 & $16.8$ \\
70083-02-05-01  &  52671.1966  &  2432  &  19.39 &  0.047 & $17.1$ \\
70083-02-06-01  &  52671.7737  &  1824  &   8.30 &  0.178 & $1.8$ \\
70083-02-06-00  &  52671.8487  &  3408  &   9.06 &  0.195 & $-2.3$ \\
70083-02-07-00  &  52672.1834  &  2480  &  16.99 &  0.271 & $-36.6$ \\
70083-02-07-01  &  52672.2507  &  1856  &  14.01 &  0.287 & $-18.4$ \\
70083-02-08-01  &  52672.6924  &  1344  &  12.90 &  0.387 & $-61.1$ \\
70083-02-08-00  &  52672.7608  &  1728  &  11.97 &  0.403 & $-54.3$ \\
70083-02-09-00  &  52673.1705  &  2560  &  27.88 &  0.496 & $-60.3$ \\
70083-02-10-02  &  52673.5744  &   944  &   9.03 &  0.587 & $-59.2$ \\
70083-02-10-01  &  52673.6796  &  1312  &  14.04 &  0.611 & $-57.3$ \\
70083-02-10-00  &  52673.7573  &  1792  &  18.22 &  0.629 & $-44.0$ \\
70083-02-11-00  &  52674.1903  &  2560  &   9.01 &  0.727 & $-13.5$ \\
70083-02-12-00  &  52674.7445  &  1680  &   6.08 &  0.853 & $25.7$ \\
70083-02-12-01  &  52674.8198  &  1872  &   9.37 &  0.870 & $18.5$ \\
70083-02-13-00  &  52675.1768  &  2688  &   5.23 &  0.952 & $41.5$ \\
70083-02-14-00  &  52675.7355  &  2112  &  17.71 &  0.079 & $38.7$ \\
70083-02-14-01  &  52675.8025  &  2528  &  11.68 &  0.094 & $37.2$ \\
70083-02-15-00  &  52676.1636  &  2688  &  10.81 &  0.176 & $19.9$ \\
70083-02-16-00  &  52676.6994  &  1120  &  11.96 &  0.298 & $-18.3$ \\
70083-02-17-00  &  52677.0792  &  3456  &  10.55 &  0.384 & $-35.3$ \\
70083-02-18-00  &  52677.6855  &  1232  &  10.86 &  0.522 & $-43.8$ \\
70083-02-19-00  &  52678.0684  &  3296  &   9.68 &  0.609 & $-33.1$ \\
\enddata
\tablenotetext{a}{Midpoint of observation, ${\rm MJD}={\rm JD}-2,400,000.5$.}
\tablenotetext{b}{Source count rate (2--30 keV) using PCU No. 2. For reference,
  1 Crab = 2500 cts s$^{-1}$.}
\tablenotetext{c}{Orbital phase corresponding to the observation
  midpoint calculated assuming $P_{\rm orb} = 4.4007$ d and the time
  of phase zero $\tau_{90} = {\rm MJD}~ 52631.383$.}
\tablenotetext{d}{Pulse arrival time delay at the Solar System
barycenter with respect to a clock with constant period $P =
604.689$~s.}
\label{tab:pulse}
\end{deluxetable*}


\begin{figure*}[t]
\begin{center}
\includegraphics[width=5.0in,angle=90]{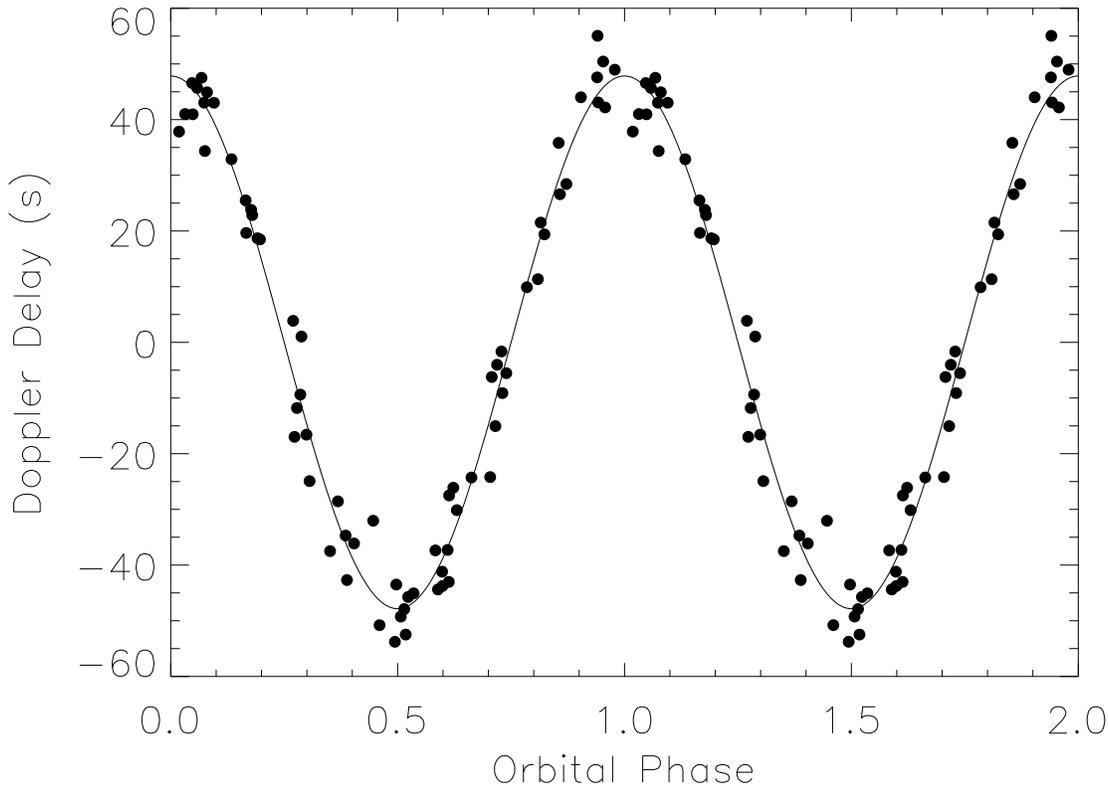}
\caption{Doppler delay curve for the same set of observations as shown
in Fig. \ref{fig:delay} but plotted as a function of orbital phase.
Each measurement is plotted twice in order to show two complete
orbital cycles.  The quadratic term representing the change in
intrinsic pulse period, evident in Fig. \ref{fig:delay}, has been
removed.  The curve represents the best-fit circular orbit.
\label{fig:delayfold}}
\end{center}
\end{figure*}

\begin{figure*}[t]
\begin{center}
\includegraphics[width=5.0in,angle=90]{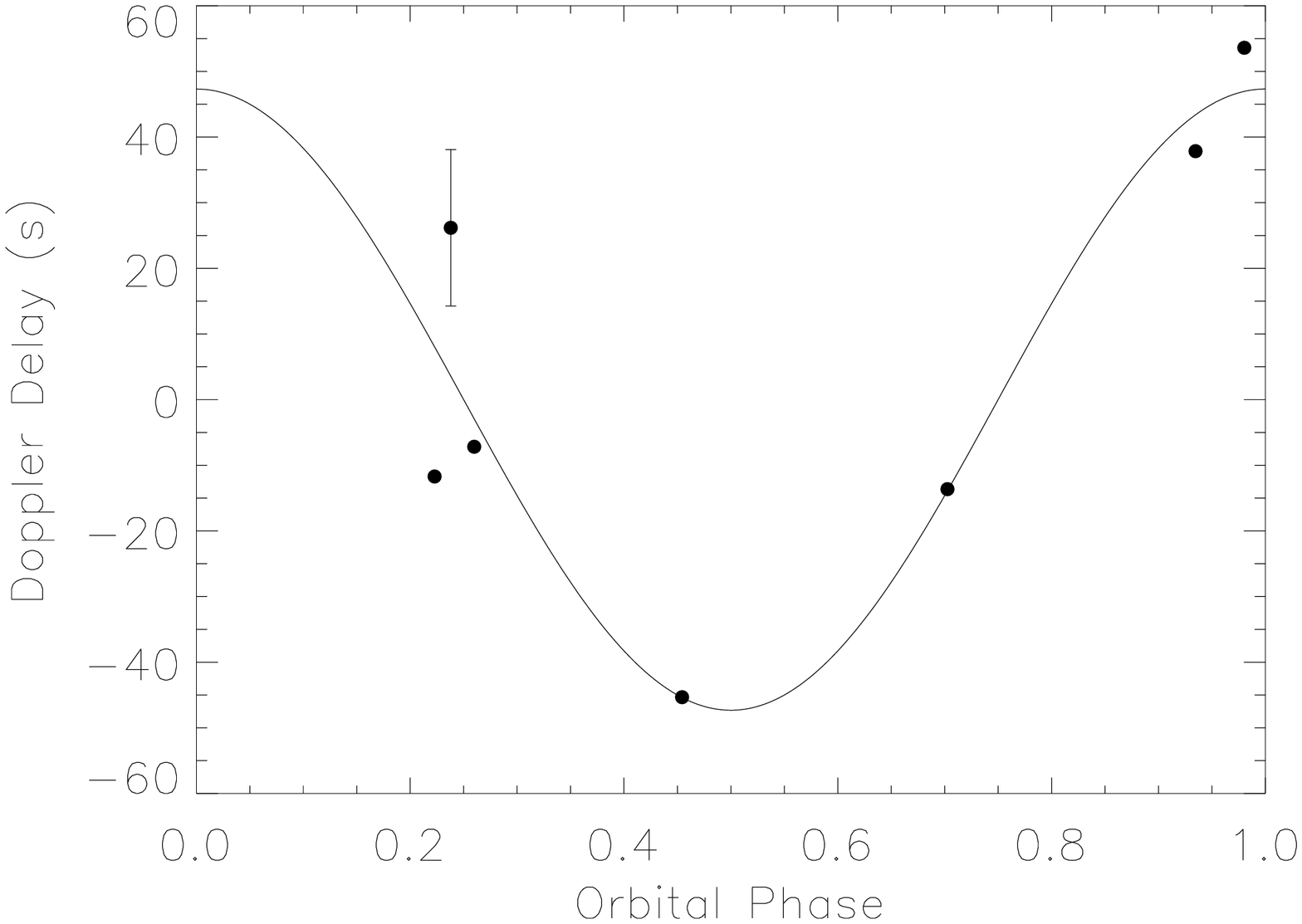}
\caption{Seven pulse arrival time delays for observations during 2000
November (approximately two years prior to the more extensive
observations shown in Fig. \ref{fig:delay}).  These data spanned only
one orbital cycle of X1908+075.  The curve represents the best-fit
circular orbit (see text).  The uncertainty in the arrival time delay
($\pm 1~\sigma$), estimated from the rms deviation of the measurements
relative to the curve, is shown as an error bar on one of the
measurements. \label{fig:delayold}}
\end{center}
\end{figure*}


The results of this fit are given in Table \ref{timeparam}.  The value
of $a_x \sin i$ of 47.8 lt-s, combined with an orbital period of 4.4
days, yields a mass function of $6.07 \pm 0.35~M_\sun$.  This points to a fairly
massive companion star.  The best-fit orbital eccentricity is $e =
0.021 \pm 0.039$, and, therefore, the orbit is consistent with being
circular.  We set a $2\sigma$ upper limit $e < 0.1$.  The pulse
period, after corrections for orbital motion of the Earth and of the
pulsar, is determined to be $P = 604.684 \pm 0.001$~s at $t = $MJD
52643.3.  The pulse period derivative is found to be positive, $\dot P
= (1.22 \pm 0.09) \times 10^{-8}$, and thus the neutron star rotation
is slowing down.  Expressed as a fractional rate, we obtain $\dot P/P
\simeq 6.4 \times 10^{-4}$ yr$^{-1}$.


\begin{deluxetable*}{cccc}
\tabletypesize{\footnotesize}
\tablecaption{Results of the Timing Analysis}
\tablehead{
\colhead{Parameter}   &  \colhead{2nd-Epoch Parameter\tablenotemark{a}}
  &  \colhead{1st-Epoch Parameter\tablenotemark{b}}  & \colhead{Comments} }
\startdata
$a_x \sin i$ (lt-s) & $47.83   \pm   0.94$ & 47.83 (fixed) &  \\
$\tau_{90} $(MJD)  & $52631.383  \pm  0.013$ & $51870.06  \pm  0.15$ & 
 superior conjunction for circular orbit  \\
$P $ (s) & $604.684 \pm  0.001$ & $604.660  \pm  0.040 $ & 
   at MJD 52643.3, 51870, respectively \\
$\dot P$ (s s$^{-1})$ & $(1.22 \pm 0.09) \times 10^{-8}$ & 0 (fixed) &  \\
$e$ & $0.021 \pm  0.039 $ & 0 (fixed) & $< 0.1$ ($2\sigma$) \\
$P_{\rm orb}$ (days) & $4.4007 \pm 0.0009$ & $ 4.4007 \pm 0.0009$ &
 determined from both epochs jointly \\
$f(M) (M_\sun)$ & $6.07 \pm 0.35$ & NA &  \\
\enddata
\label{timeparam}
\tablenotetext{a}{Based on 69 pulse arrival times obtained from 2002
December 23 through 2003 February 8.  Errors cited are
single-parameter $1 \sigma$ confidence limits.}
\tablenotetext{b}{Based on 7 pulse arrival times obtained from 2000
November 19 through 2000 November 23. Errors cited are
single-parameter $1 \sigma$ confidence limits.}

\end{deluxetable*}


In Fig. \ref{fig:delayfold} we show the measured pulse arrival time
delays for the extensive 2002--2003 epoch 2 data set relative to our
best-fit model excluding the orbital Doppler shift terms.  The results
are plotted modulo the 4.4-day orbital period.  The solid curve is the
best fit circular orbit with $a_x \sin i = 47.8$ lt-s.

A similar pulse phase analysis was then carried out on the smaller
epoch 1 data set taken approximately two years earlier.  Only 7 of the
pulse arrival time delays from this earlier set proved to be useful in
the model fitting, in part because it was difficult to unambiguously
connect the pulse phase across the large gap between the two subsets
of epoch 1 observations.  The Doppler delays for these 7 arrival
times, which span only one orbital cycle, are plotted in Fig,
\ref{fig:delayold}.  When fitting these delays, we fixed the amplitude
($a_x \sin i$) at the value determined from the later, more extensive
observations, which yield much more accurate orbital parameters.  The
orbital period ($P_{\rm orb}$) was fixed at the value determined
previously from 3 years of ASM observations (WRB).  The main objective
in using these observations was to determine a relatively accurate
orbital phase some two years before the later, more extensive
observations.  The two orbital phase determinations, separated by
$\sim 2$ years, are then combined to compute a more accurate
determination of the orbital period, $P_{\rm orb} = 4.4007 \pm 0.0009$
days.  This is consistent with, and of comparable precision to, the
orbital period of $4.400 \pm 0.001$ days determined from the ASM X-ray
light curve by WRB.

By now (2004 April), the ASM has accumulated $\sim 40,000$ individual
intensity measurements of X1908+075 over $\sim 8$ years.  This is more
than twice the amount of data contained in the original light curve of
WRB.  We have therefore redone the light curve analysis using all the
presently available ASM data to derive a new X-ray intensity-based
value for the orbital period.  Our result is $P_{\rm orb} = 4.4006 \pm
0.0006$ days.  This is consistent with the earlier result of WRB and
with our pulse-timing-based result as well.  The folded ASM light
curves are shown in Fig. \ref{fig:asmfold}.

\section{ORBITAL PHASE-DEPENDENT SPECTRAL ANALYSIS}

\noindent

\begin{figure}[t]
\begin{center}
\includegraphics[width=3.2in,angle=0]{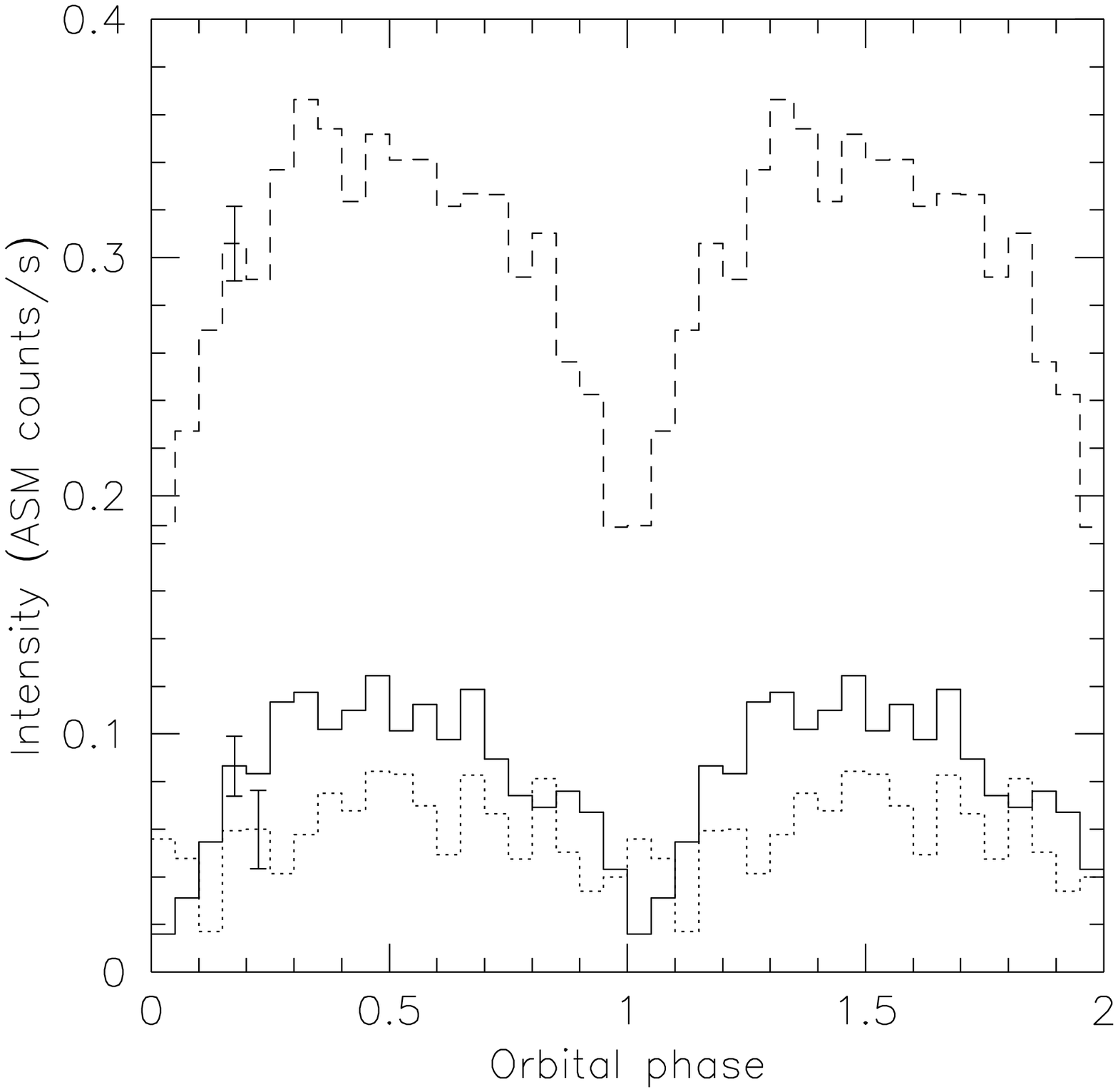}
\caption{Average intensity as a function of orbital phase produced by
folding $\sim 8$ years of ASM results at the period $P = 4.4006$
days. The dotted, solid, and dashed histograms correspond to the
(nominal) 1-3.5, 3.5-5, and 5-12 keV energy bands.  Each curve is
shown for two cycles.  For each of these energy bands, an intensity of
1 ASM count s$^{-1}$ corresponds to $\sim 40$ mCrab.
\label{fig:asmfold}}
\end{center}
\end{figure}

The X-ray intensity of X1908+075 is strongly modulated with orbital
phase.  A qualitative feeling for the energy-dependence of the
modulation can be obtained from the folded ASM data
(Fig. \ref{fig:asmfold}).  The ASM light curves are extremely well
sampled both as a function of orbital phase and over many orbital
cycles.  We define the degree of modulation as $(I_{\rm max}-I_{\rm
min})/(I_{\rm max}+I_{\rm min})$, where $I_{\rm max}$ and $I_{\rm
min}$ are the maximum and minimum intensities observed in the light
curves.  The degrees of modulation are found to be $0.41 \pm 0.18$,
$0.70 \pm 0.15$, and $0.38 \pm 0.04$ for the $1.5-3$, $3-5$, and
$5-12$ keV energy bands, respectively.  The uncertainties have been
estimated primarily on the basis of statistical fluctuations in
$I_{\rm min}$, and do not include any contributions due to possible
systematic baseline offsets.  Nonetheless, we take these numbers, as
well as the different shapes of the folded light curves, to be
evidence of energy dependence of the orbital modulation.


In spite of the fact that the orbital light curves produced with the
ASM are very well sampled, the energy resolution (limited to 3
channels) and calibration are not good for determining detailed
spectral parameters.  The PCA data are better suited to this purpose.

To help guide the spectral analysis, we consider the overall average
spectrum of X1908+075 during the epoch 1 and epoch 2 {\it RXTE}
pointed observations. In each case, the PCA spectrum (2-25 keV; PCU 2)
shows evidence of low-energy attenuation by a substantial column
density and the presence of line emission due to neutral or ionized Fe
at energies of approximately 6.4-6.7 keV.  The HEXTE spectrum (both
clusters) in the range 15-70 keV shows a high-energy cutoff that rules
out a simple power-law model.  For both epochs, thermal bremsstrahlung
and cutoff power-law models yield reasonable fits of the continuum
spectrum.  We proceed with the bremsstrahlung model since it has fewer
free parameters, gives results for each epoch that are remarkably
consistent, and (in hindsight) gives best-fit temperatures from the
individual observations that agree well with each other.  Since the
spectral resolution of the PCA is at best marginal for determination
of the Fe line energy, we assume that any line emission near these
energies is at 6.7 keV; this should be an adequate model of the Fe
line region for the present purposes.  The fit of the average spectrum
from epoch 1 (35 ks of exposure) yields a column density $N_H = (1.12
\pm 0.02) \times 10^{23}$ cm$^{-2}$, a temperature $kT = 23.3 \pm 0.5$
keV, and a value of chi-square per degree of freedom $\chi^2_{\nu}$ =
2.42.  The fit of the spectrum from 153 ks of exposure in epoch 2
yields $N_H = (1.22 \pm 0.02) \times 10^{23}$ cm$^{-2}$, $kT = 22.6
\pm 0.3$ keV, and $\chi^2_{\nu}$ = 4.84.  These values for
$\chi^2_{\nu}$ are formally unacceptable. However, the average spectra
comprised data obtained from large accumulated exposure times so that
the uncertainties from counting statistics are very small.  In fact,
in our analysis of the PCA spectral data we estimated the error for
each energy bin by adding the statistical error in quadrature with a
presumed 1\% systematic error which represents the uncertainty in the
instrument response function \citep[e.g.,][]{sob00}.  For much of the
energy range these presumed systematic errors dominate the statistical
errors.  Perhaps most importantly with regard to the $\chi^2$ values
of the fits of the average spectra, the $N_H$ values vary
significantly from observation to observation.  In such a case, a
model based on a single value for $N_H$ cannot be expected to
precisely fit the average spectrum.

Next, we analyzed the net spectrum for each of the 69 epoch 2 pointed
observations for which we have determined a pulse arrival time. The
analysis was repeated for each of three spectral models in order to
estimate how the formal statistical uncertainties compare with those
due to the particular choice of model.  Each model is intended to
provide a simple spectral shape that is likely to fit the data well
and to yield a good estimate of the low-energy absorption.  All the
models involved a thermal bremsstrahlung spectrum, a 6.7 keV iron line
(which, as noted above, also represents line emission near 6.4 keV),
and absorption by the interstellar medium (ISM) and a stellar wind.

Inspection of the raw pulse height spectra reveals that there are
`excess' counts at low energy (i.e., $\lesssim 6$ keV) with respect to
the energy dependence expected from photoelectric absorption in a
neutral gas.  Two possible explanations of this form of the low-energy
portion of the spectra immediately arise.  First, the observed
spectrum could be formed by both X-rays that propagate directly to the
detectors along paths that pass close to the companion star and
therefore are heavily attenuated, as well as X-rays that are scattered
farther out in the stellar wind and are not attenuated so heavily by
photoelectric absorption.  Second, if the stellar wind is partially
ionized, then the energy dependence of any absorption will not be as
strong as that for strictly neutral gas.  The combination of these
effects can be approximately taken into account by allowing the
presence of two spectral components that are attenuated by different
amounts.

We take the intrinsic emission spectrum of the source to be
\begin{equation}
S(E) = C_1 Br(E,kT) + C_2 G(E)
\end{equation}
where $Br(E,kT)$ represents an optically thin thermal bremsstrahlung
spectrum characterized by temperature $T$ and $G(E)$ represents a
Gaussian profile line centered at 6.7 keV with width $\Delta E (1
\sigma) = 0.3$ keV.
This is used, in turn, in a generic spectral model:
\begin{eqnarray}
F_x & = & \exp[-N_{H_{\rm ISM}}\sigma (E)] \times \nonumber \\
 & & \{C_3 \exp[-N_{H_{\rm wind}}\sigma (E)] + (1 - C_3) \} S(E)~. 
\end{eqnarray}
$C_1$, $C_2$, and $C_3$ are parameters of the model.  Each of the
three spectral models that we apply to the observations uses a
particular form of this generic model.  In Model 1, the thermal
bremsstrahlung spectrum is attenuated by the ISM and a stellar wind,
$kT$ is free to vary from one observation to another, and $C_3$ is
assigned a value of one.  The ISM and wind column densities are not
determined separately; only their sum is determined.  Model 2 is the
same as Model 1, except that $kT$ is fixed at the value of 22.6 keV
that was determined by the fit of the average spectrum of the epoch 2
observations.  Model 3 is similar to Model 2, except that the
absorption is characterized by `partial covering' and so the parameter
$C_3$ is determined for each observation.  For Models 1 and 2, the
free parameters were simply determined by varying the free parameters
so as to minimize the $\chi^2$ statistic.  For Model 3 such a
procedure did not yield acceptable results, so an iterative procedure
was used.  In this procedure, the value of $N_{H_{\rm ISM}}$ was fixed
and the remaining free parameters including $N_{H_{\rm wind}}$ were
determined for each observation by a minimum $\chi^2$ fit with XSPEC.
A fit of a model of the orbital phase dependence of $N_{H_{\rm wind}}$
was then made to the resulting $N_{H_{\rm wind}}$ values to determine
wind parameters (with $\beta = 0$, see \S 5 below).  Then the spectral
fit of each observation was redone using XSPEC with $N_{H_{\rm ISM}}$
as a free parameter and with the value of $N_{H_{\rm wind}}$ fixed to
the value predicted from the fit of the wind parameters.  The weighted
mean of the $N_{H_{\rm ISM}}$ values then was used as a revised
estimate of this quantity.  This entire procedure was iterated 4 times
to obtain convergence.  In the final iteration, $N_{H_{\rm ISM}}$ was
set to $4.6 \times 10^{22}$ atoms cm$^{-2}$, which indicates that
substantial absorption occurs in the ISM (although it should be noted
that this column density is, at all orbital phases, comparable or
smaller than that in the stellar wind).

The statistics of the individual observations were not sufficient to
conclude that one of the spectral models fit the data significantly
better than the other models.  Rather, all three models yield
comparable quality fits with reduced $\chi^2$ values in the range
0.5-2.0 for all but a few observations.  As might be expected, the
value of the normalization constant $C_1$ does not appear to be
dependent on orbital phase, but the iron line intensity does appear to
depend on orbital phase by as much as 30\% in the Model 2 results and
less in the Model 1 and Model 3 results.  In the Model 3 fit results,
the value of $C_3$, the covering fraction, is always in the range
0.4-1.0 and, for $\sim 85$\% of the observations, is in the range
0.85-1.0.

\begin{figure*}[t]
\begin{center}
\includegraphics[width=5.0in]{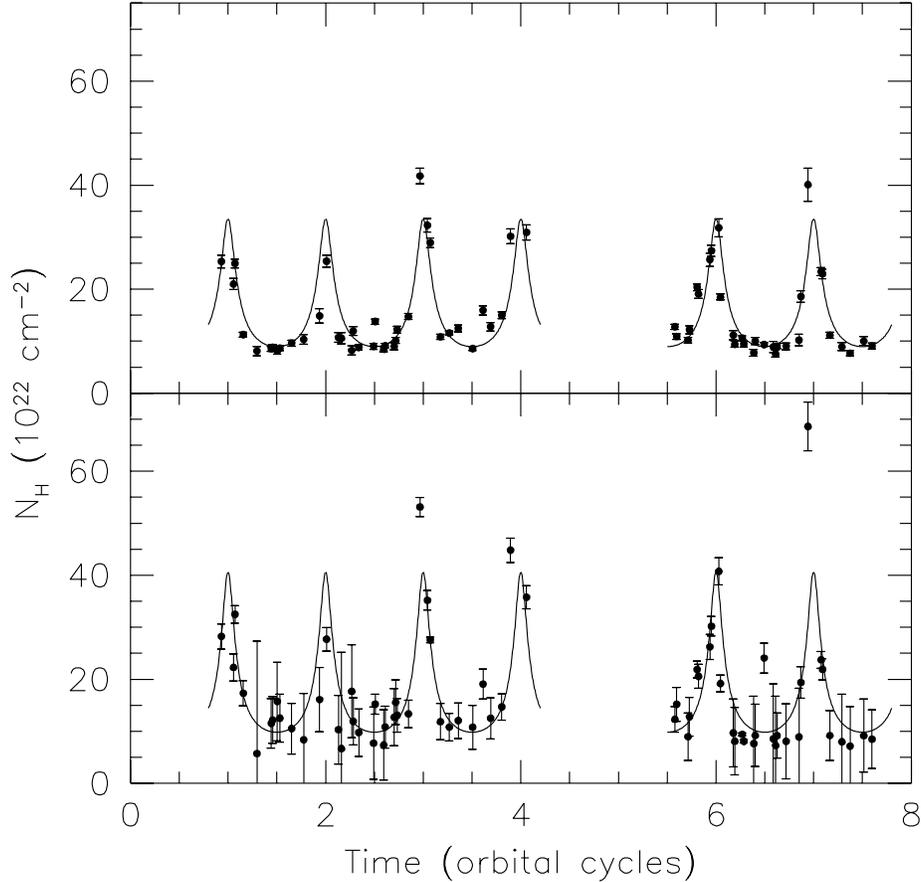}
\caption{Estimates of $N_H = N_{H_{\rm wind}} + N_{H_{\rm ISM}}$
derived from spectral fits plotted as a function of time expressed in
units of orbital cycles.  Integral values of time, i.e., the times of
orbital phase zero, correspond to superior conjunction of the neutron
star.  The upper plot shows estimates derived from fits of spectral
Model 1 with uncertainties ($\pm 1$ $\sigma$) taken from the XSPEC
fits.  The lower plot shows estimates from spectral Model 3.with
uncertainties ($\pm 1$ $\sigma$) taken from those for $N_{H_{\rm
wind}}$ from the final set of XSPEC runs since we assume the
uncertainty in $N_{H_{\rm ISM}}$ may be neglected in this case.  The
curves show the best fits of the orbital-phase dependent column
density with $\beta = 0$; the best-fit curve for $\beta = 1$ (see
text) is nearly indistinguishable from the curve shown here.  For this
plot, the long gap between sets of observations was artificially
shortened by 4.0 orbital cycles.
\label{fig:NHcurve}}
\end{center}
\end{figure*}

\begin{figure*}[t]
\begin{center}
\includegraphics[width=5.0in,angle=0]{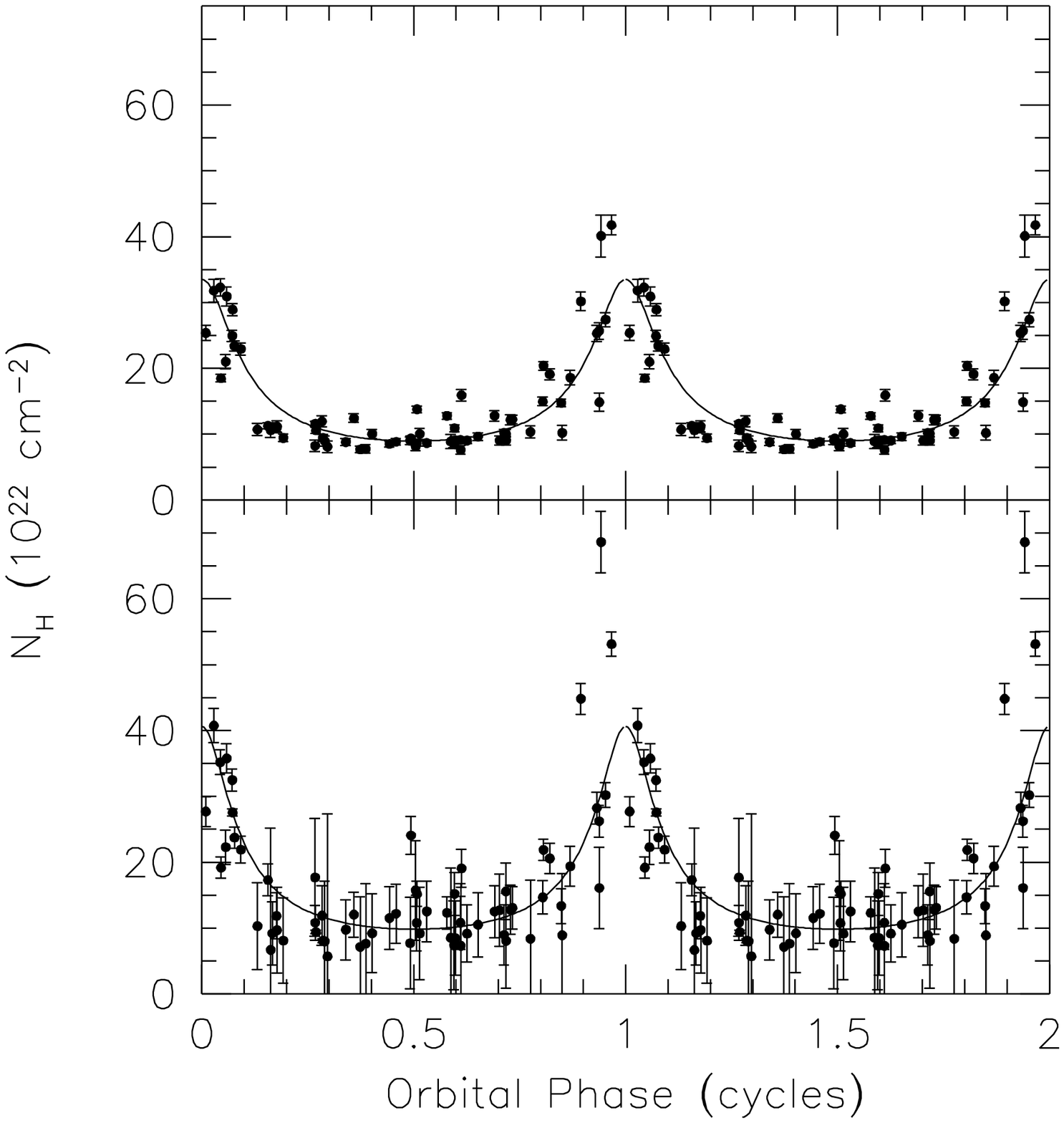}
\caption{(top) $N_H$ estimates for spectral Model 1 as a function of
orbital phase.  The points shown are those shown in the top panel of
Fig. \ref{fig:NHcurve}. (bottom) $N_H$ estimates for spectral Model 3.
In both panels, the solid curve shows the best-fit $\beta = 0$ stellar
wind model.  As noted in Fig.\ref{fig:NHcurve}, the best-fit curve for
the $\beta = 1$ model is nearly indistinguishable. The points and
curves are shown in each of two cycles.
\label{fig:NHfoldnoav}}
\end{center}
\end{figure*}

The total neutral hydrogen column densities ($N_H = N_{H_{\rm ISM}} +
N_{H_{\rm wind}}$) for the 69 observations are shown as a function of
time in Fig. \ref{fig:NHcurve}; the top panel shows results from
spectral Model 1, while the bottom panel shows those for the partial
covering model (Model 3).  Note the pronounced variation of $N_H$ with
maxima at the expected orbital phases.  A comparison of the two panels
provides an idea of the sensitivity of $N_H$ to the choice of spectral
model; spectral Model 2 yields qualitatively similar results.  The
fitted values of $N_H$ based on Models 1 and 3 are also plotted modulo
the orbital phase in Figure \ref{fig:NHfoldnoav}.


\begin{figure}[t]
\begin{center}
\includegraphics[width=3.5in]{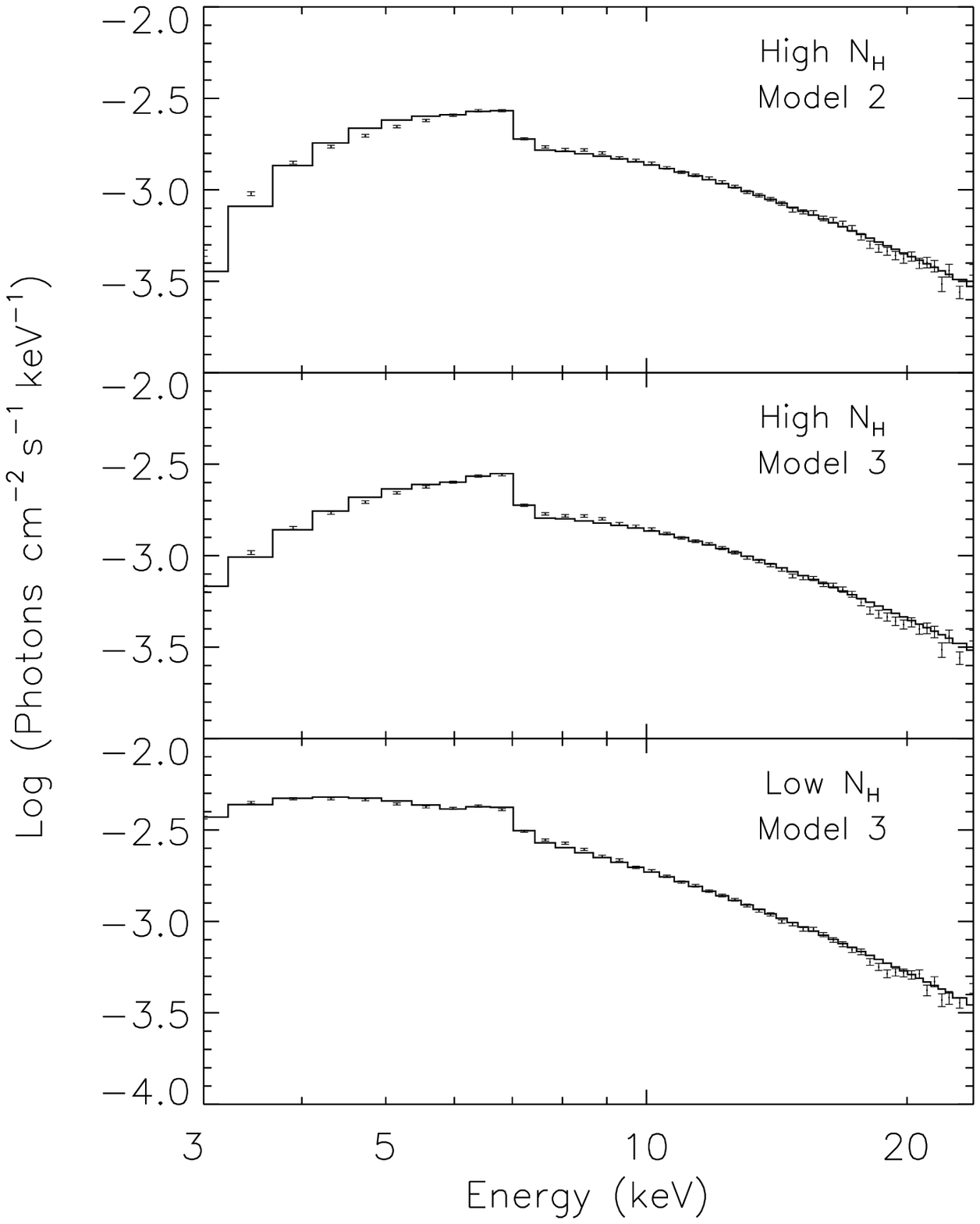}
\caption{Inferred photon number spectra (points with error bars) of the
accumulated data from 7 observations when the absorbing column density
was inferred to be relatively high (top and middle panels), and of the
data from 8 observations when the column density was relatively low
(bottom panel).  The histogram in the top panel shows the best-fit of
the Model 2 spectrum while the histograms in the middle and lower
panels show the best fit of the Model 3 spectrum.
\label{fig:spectra}}
\end{center}
\end{figure}

The average spectrum from seven observations performed at orbital
phases when the column density is expected to be high, i.e.,
$\vert\phi_{orb}\vert < 0.12$, is shown in the top and middle panels
of Figure \ref{fig:spectra}. The degree of low energy attenuation
contrasts with that of the spectrum from eight observations performed
at orbital phases when the column density is expected to be low
(Fig. \ref{fig:spectra}, lower panel).  We have fit each average
spectrum with both Models 2 and 3, but we fixed the value of the
bremsstrahlung temperature in each case to that determined from the
average spectrum of all the epoch 2 observations.  We also fixed the
energy of the Gaussian line center at 6.7 keV and the line width at
$\sigma_{line} = 0.3$ keV.  The results are given in Table
\ref{sp-param}.  The Model 2 best-fit high $N_H$ spectrum and the
Model 3 best-fit low and high $N_H$ spectra are also shown in
Fig. \ref{fig:spectra}.  The plots illustrate the low energy
attenuation and its variation.  Spectral features are quite apparent
at $\sim 7$ keV; these comprise Fe K line emission and absorption
edges.  The iron absorption is particularly prominent in the high
$N_H$ spectrum.  Both the low and high $N_H$ spectra are fit somewhat
better by Model 3, although neither model fits the data well; the
values of reduced $\chi^2$ are significantly above 1.

From the spectral fits, we find that the X-ray flux, averaged over an
observation and adjusted for low-energy absorption, is most often in
the range $F_X$(2--30 keV) $\sim$ 3--8 $\times 10^{-10}$ ergs
cm$^{-2}$ s$^{-1}$.  Thus the X-ray luminosity of X1908+075 is
typically $L_X$(2--30 keV) $\sim$ 2--6 $\times 10^{36}\ (D/8\ {\rm
kpc})^2$ ergs s$^{-1}$ where $D$ is the distance to the source, and,
even at the high end of this range, is far below the Eddington limit
for a 1.4 $M_\sun$ neutron star, i.e., $L_{Edd} = 2.0 \times 10^{38}$
ergs s$^{-1}$.  This is consistent with the picture that the neutron
star is accreting from a wind from the companion star rather than
being fed via Roche lobe overflow.


\begin{deluxetable*}{lcccc}
\tablecaption{Spectral Fit Results}
\tablehead{
 & \colhead{Low $N_H$} & \colhead{Low $N_H$} & \colhead{High $N_H$}
 & \colhead{High $N_H$} \\ 
\colhead{Parameter} & \colhead{Model 2} & \colhead{Model 3} 
 & \colhead{Model 2} & \colhead{Model 3}
 }
\startdata
$C_1$\tablenotemark{a} & $0.1012 \pm 0.0004$ & $0.1023 \pm 0.0005$ &
  $0.0880 \pm  0.0004$  & $0.0905 \pm 0.0005$ \\
$C_2$\tablenotemark{b} &  $5.7 \pm 1.1$ & $6.1 \pm 1.1$ &
  $3.3 \pm 1.0$ & $4.8 \pm 1.1$ \\
$N_{H_{\rm ISM}}$\tablenotemark{c} & & 4.7 (fixed) & & 4.7 (fixed) \\
$N_{H_{\rm wind}}$\tablenotemark{c} & & $8.5 \pm 1.3$ & &  $24.7 \pm 0.7$ \\
$N_H$(total)\tablenotemark{c} & $9.3 \pm 0.1$ & & $24.3 \pm 0.2$ & \\
$C_3$ (covering fraction)& & $0.66 \pm 0.06$ & & $0.916 \pm 0.008$ \\
$\chi^2$ & 126.2 & 115.0 & 222.3 & 142.5 \\
Degrees of freedom & 46 & 45 & 46 & 45 \\
$\chi^2_\nu$ & 2.74 & 2.56 & 4.83 & 3.17 \\
\enddata
\label{sp-param}
\tablenotetext{a}{Normalization factor for XSPEC component ``bremss''
to give flux in units of photons cm$^{-2}$ s$^{-1}$ keV$^{-1}$}
\tablenotetext{b}{Iron line flux in units of $10^{-4}$
photons cm$^{-2}$ s$^{-1}$} 
\tablenotetext{c}{In units of $10^{22}$ H-atoms cm$^{-2}$}
\tablecomments{All quantities are defined in eqs. 2 and 3.  For
all 4 fits, the bremsstrahlung temperature was fixed at $kT = 22.6$
keV, the spectral line center energy was fixed at 6.7 keV and the line
width fixed at $\sigma = 0.3$ keV.  The low $N_H$ spectrum is the
average from 8 observations.  The high $N_H$ spectrum is the average
from 7 observations (see text).}
\end{deluxetable*}


\section{ABSORPTION IN A STELLAR WIND AND THE BINARY SYSTEM}

The variation of the column density with orbital phase is caused by
the movement of the X-ray source through the stellar wind of the
companion star which is likely a spatially and temporally complex
medium.  In order to estimate the orbital inclination as well as other
orbital parameters and properties of the companion star, we proceed by
modelling the column density variation as orbital-phase dependent
absorption in a spherically symmetric and temporally constant wind
from the companion star.

The wind from an early-type star is often modelled as a steady-state,
spherically symmetric flow at a velocity that gradually increases from
zero to a terminal velocity of order of magnitude $\sim 1000$ km
s$^{-1}$. The wind density profile then follows from the assumption of a
radius-independent mass flux \citep[and references
therein]{lucy70,castor75,iswbook99}.  For our purposes the radial flow
velocity may be taken to be
\begin{equation}
v(r) = v_{\infty}(1 - R_c/r)^{\beta}
\end{equation}
where $r$ is the distance from the center of the companion star,
$v_{\infty}$ is the wind terminal velocity, $R_c$ is the radius of the
companion star, and $\beta$ is often in the range 0.7--1.2 for
early-type stars \citep[e.g.,][]{gl89,puls96}.  The wind density
profile then follows from the conservation of mass:
\begin{equation}
n(r) = \frac{n_0(1 - R_c/a)^{\beta}}{(r/a)^2(1 - R_c/r)^{\beta}}
\end{equation}
where $n(r)$ is the number density of hydrogen atoms in the stellar
wind at the distance $r$ from the companion star and $n_0$ is the
number density at the distance $a$.  For a circular orbit with orbital
inclination angle $i$, the instantaneous column density of material
between the neutron star and the observer, $N_H$, is given by:
\begin{eqnarray}
N_H & = & N_{H_{\rm ISM}} + N_{H_{\rm wind}} = N_{H_{\rm ISM}} + \int_{0}^{\infty}n[r(s)]ds\\
 & = & N_{H_{\rm ISM}} + a n_0 (1 - R_c/a)^{\beta}
\int_{0}^{\infty}\frac{ds^\prime}{r^{\prime 2}(1 - R_c^\prime/r^\prime)^{\beta}}
\end{eqnarray}
where $N_{H_{\rm ISM}}$ and $N_{H_{\rm wind}}$ are the separate
contributions from the interstellar medium and the stellar wind,
respectively, $s$ is the distance along the line from the neutron star
toward the observer, and primed quantities are normalized relative to
$a$.

The radial distance $r$ is a function of the distance $s$ and the
angle $\phi$ subtended at the neutron star between the direction to
the observer and the radial direction.  In terms of normalized
quantities, we have
\begin{equation}
r^{\prime 2} = 1 + s^{\prime 2} - 2 s^{\prime} \cos \phi~.
\end{equation}
The angle $\phi$, in turn, is related to the inclination angle and the
orbital phase by:
\begin{equation}
\cos \phi = -\sin i \cos[\Omega (t-\tau_{90})]~.
\end{equation}
For the simple case of an inverse square law density profile, i.e.,
$\beta = 0$, the wind column density integral equals $a n_0 \phi/\sin
\phi$ if $\phi$ is given in radians.  In this case, the wind column
densities, in units of $n_0 a$, are $1$, $\pi/2$, $3\pi \sqrt{2}/4$,
and $\infty$ for $\phi$ = 0, $\pi/2$, $3\pi/4$, and $\pi$ radians
respectively, and the ratio of maximum to minimum value is given
simply by $N_{H_{\rm wind}}{\rm (max)}/N_{H_{\rm wind}}{\rm (min)} =
(\pi/2+i)/(\pi/2-i)$.

Since we know $\tau_{90}$ from our orbital analysis, there are 5
parameters that can potentially be determined from fits to the orbital
modulation of $N_H$: $i$, $\beta$, $R_c$, $N_{H_{\rm ISM}}$, and the
product $a n_0 (1 - R_c/a)^{\beta}$.  In the analysis procedure, for
each of a set of assumed trial values for $i$, $\beta$, and $R_c$, we
performed a linear least squares fit of the model defined by eq. (7)
to the column densities determined from our Model 1 spectral fits, and
thereby obtained values for $N_{H_{\rm ISM}}$ and $a n_0 (1 -
R_c/a)^{\beta}$ as well as an estimate of the root-mean-square
deviation of the column densities from the fit.  We have fit models
for a wide range of orbital inclinations.  For each trial value of
$i$, we use our best-fit value for $f(M)$ (see Table \ref{timeparam})
and assume that the mass of the neutron star is the canonical $1.4
M_\sun$ to compute the mass of the companion $M_c$.  Using the
resulting mass ratio, $i$, and our best fit value for $a_x \sin i$, we
then compute the orbital separation $a$.  For the given inclination,
the set of trial values of $R_c$ covers a range up to a maximum value
taken to be the smaller of the Roche lobe radius of the companion star
or the radius which would produce a grazing eclipse.  Finally, for
simplicity, we considered only three discrete values for the wind
parameter $\beta$: 0, 1/2, and 1.

\begin{figure*}[t]
\begin{center}
\includegraphics[width=5.0in]{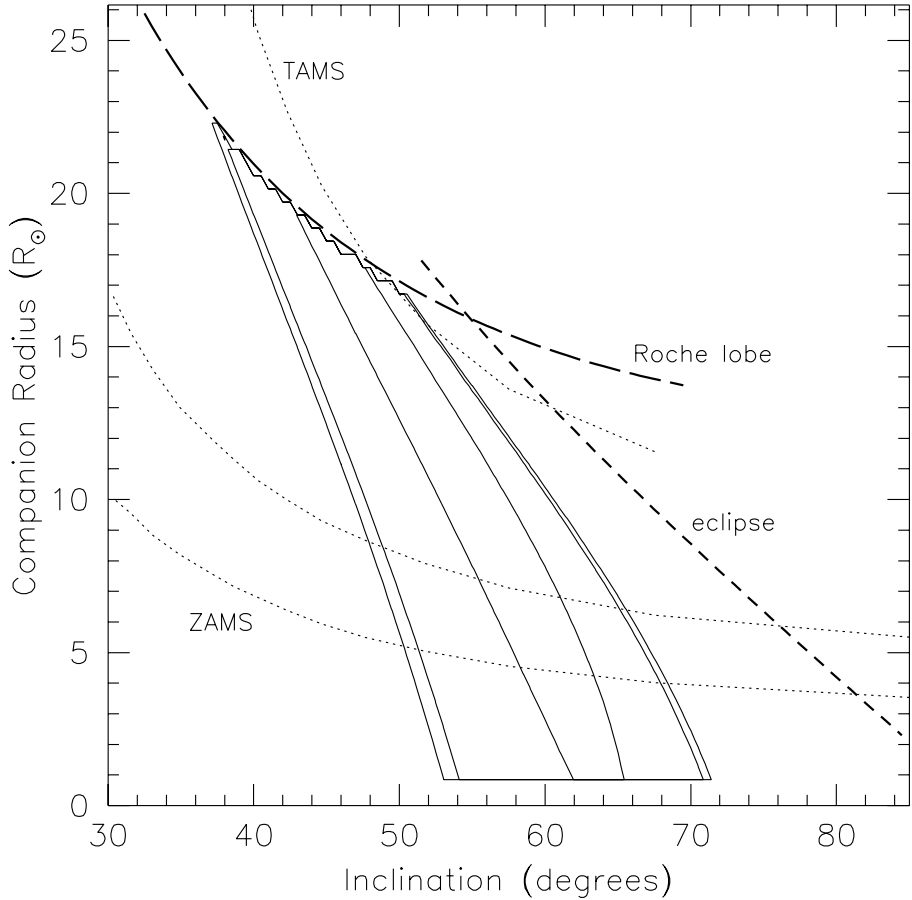}

\caption{Contours (solid curves) of $\chi^2$ in the $i-R_c$ plane from
fits of the column densities for wind density profiles with $\beta=1$
to the spectral Model 1 $N_H$ measurements shown in the upper panel of
Fig. \ref{fig:NHcurve}. The three contours show the values for which
$\chi^2 = ( 1 + \Delta / 65) \chi^2_{min}$ where $\Delta =$ 0.25, 4.0,
or 4.61.  The outer contour corresponds to the formal 90\% confidence
limit for two interesting parameters.  For each value of the
inclination, the mass of the companion star is determined to within a
small uncertainty.  The long-dash curve represents the radius of the
Roche lobe of the companion star, the medium-dash curve represents the
radii where grazing eclipses would occur, and the three short-dash
curves represent the radii of stars on the zero-age main sequence
(ZAMS), stars that have exhausted half of the hydrogen at their
centers, and stars on the the terminal age main sequence (TAMS; see
text).  The mass-radius relations were provided by Podsiadlowski
(2004, private communication).\label{fig:chisqr}}

\end{center}
\end{figure*}

Since we do not have any reliable way to estimate the uncertainties in
the measured column densities due to systematic errors, we have taken
the rms spread between the values found from the spectral analysis
(for the 69 individual observations) and the best-fit model to be the
$1 \sigma$ error.  When we do this, the minimum value of $\chi^2$
should be 65, by definition, since there are 69 data points and 4
fitted parameters (not counting $\beta$). The contour at $\chi^2 \sim
65+4.6 = 69.6$ is then relevant for estimation of the 90\% confidence
region in the $i - R_c$ plane (with $N_{H_{\rm ISM}}$ and $a n_0 (1 -
R_c/a)^{\beta}$ treated as uninteresting parameters).  In
Fig. \ref{fig:chisqr}, we show contours of $\chi^2$ in this plane from
fits of the $\beta = 1$ model.  We note that, in the limit of
vanishing $R_c$, the value of $\beta$ becomes irrelevant and the model
is equivalent to the $\beta = 0$ model.  Thus, the contours at small
$R_c$ in Fig. \ref{fig:chisqr} indicate the acceptable range of
inclinations in the $\beta = 0$ fits.  We find that the best fits for
each of the three values of $\beta$ are very similar in quality, so we
cannot discriminate among these values of $\beta$ using this criterion
alone.

The best fits of the $\beta = 1$ model to the values of $N_H$ from
spectral Models 1, 2, and 3 give rms observed--minus--calculated
values of $4.1 \times 10^{22}$, $5.1 \times 10^{22}$, and $7.7 \times
10^{22}$ H atoms cm$^{-2}$, respectively.  The $\beta =$ 0 and 1/2
cases give slightly larger residuals.  The best fit to the spectral
Model 1 results of a $\beta =$ 0 model is characterized by $i =
64.5\arcdeg$, $M_c = 11 M_\sun$, $a = 60$ lt-s, $N_{H_{\rm ISM}} = 4.1
\times 10^{22}$ cm$^{-2}$, and $N_{H_{\rm wind}}(\phi=0) = a n_0 = 4.7
\times 10^{22}$ cm$^{-2}$.  The best fit to the spectral Model 1
results of a $\beta =$ 1 model has $i = 48.5\arcdeg$, $M_c = 17
M_\sun$, $R_c = 16 R_\sun$, $a = 69$ lt-s, $N_{H_{\rm ISM}} = 5.0
\times 10^{22}$ cm$^{-2}$, and $N_{H_{\rm wind}}(\phi=0) = 3.6 \times
10^{22}$ cm$^{-2}$.  The column densities predicted by these models
are practically indistinguishable from each other; they are shown in
Figs. \ref{fig:NHcurve} and \ref{fig:NHfoldnoav}. We emphasize that
acceptable fits are obtained for a considerable range of parameters,
and that the best-fit parameters given above are merely illustrative
of the allowed values.  Our estimates of the system parameters for
X1908+075 are summarized in Table \ref{sysparam}.

 
\begin{figure*}[t]
\begin{center}
\includegraphics[width=5.0in]{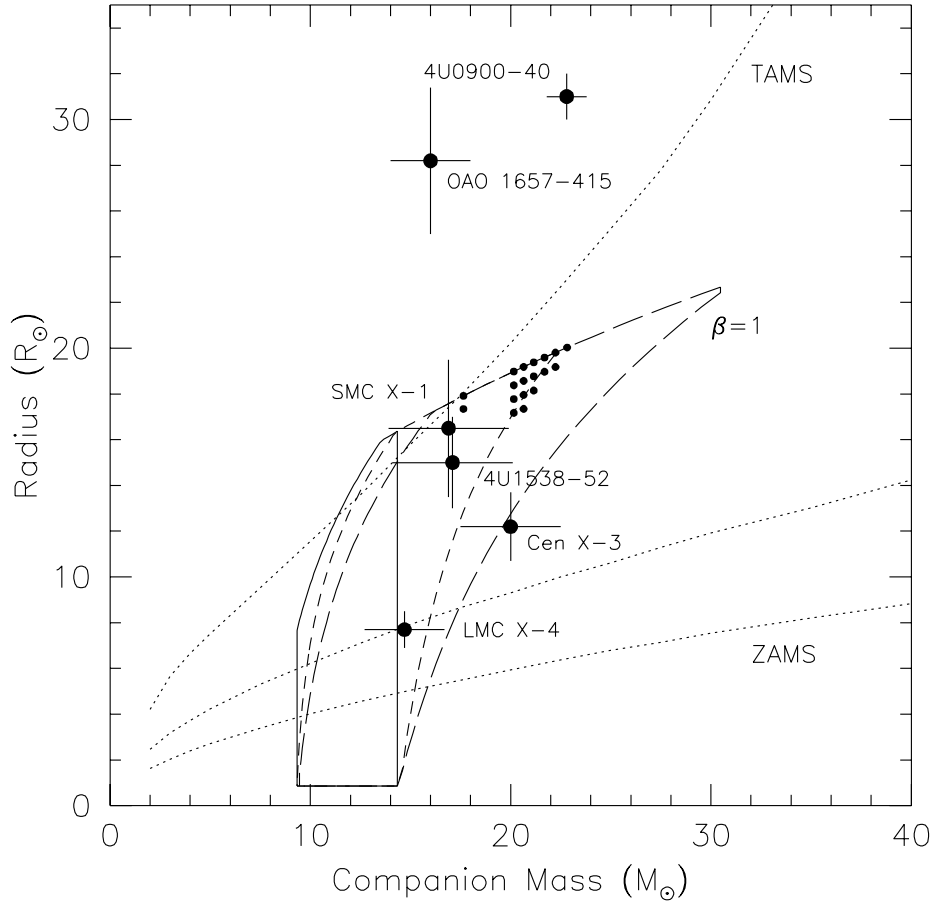}

\caption{Limits on the companion star radius as a function of the
companion mass determined from fits of the variation of column density
with orbital phase.  The solid, short-dashed, and long-dashed lines
show limits from the $\beta = $0, 1/2, and 1 fits, respectively, and
correspond to the formal 90\% confidence contours (cf.
Fig. \ref{fig:chisqr}).  High values of $M_c$ correspond to low values
of $i$ and vice versa.  The lower bounds on the radius are arbitrarily
set at $\sim 1\ R_\sun$.  The medium-size dots show those fits for
which the wind mass loss rate inferred from the observations is no
more than a factor of 3 greater than the rate predicted by the recipe
in \citet{vink00}, (see text).  The dotted curves are mass-radius
relations for main-sequence stars where the mass is that indicated on
the horizontal axis; the lowest curve represents the zero-age main
sequence, the middle curve represents stars that have exhausted half
of the hydrogen at their centers, and the upper curve represents the
terminal age main sequence (see text).  The mass-radius relations were
provided by Podsiadlowski (2004, private communication).  Also shown
on the plot are the measured masses and radii of the ``normal'' stars
in six well known high-mass X-ray binaries (\citealt{jr84,nag89};
A. Levine 1984, private communication).  The associated error bars
crudely represent 90\% confidence uncertainties.
\label{fig:massfunc}}

\end{center}
\end{figure*}

The results of the column density fits are shown as constraints on the
companion star mass and radius in Fig. \ref{fig:massfunc}. and allow
us to set, e.g., assuming $\beta = 1.0$, upper limits of $\sim 30
M_\sun$ and $\sim 22 R_\sun$ on the mass and radius, respectively, of
the companion star.  We note that for $M_c > 13.5 M_\sun$, the upper
limits on the companion radius correspond to the Roche lobe radius,
while the lack of an apparent X-ray eclipse bounds the allowed radii
for $M_c < 13.5 M_\sun$.  In Fig. \ref{fig:massfunc} we also show the
measured masses and radii of the companion stars in 6 well known
high-mass X-ray binaries and mass-radius relations for zero-age
main-sequence (ZAMS) stars, for stars that have exhausted half of the
hydrogen at their centers, and for stars at the end of the main
sequence phase of their evolution, i.e., the terminal-age main
sequence (TAMS), where the mass is that indicated on the axis
(Ph. Podsiadlowski 2004, private communication).  These mass-radius
relations were derived for single (isolated) stars neglecting the
effects of any mass loss.

Each set of system parameters and wind density profile that is
consistent with the column density measurements yields a wind mass
flux if we assume a wind terminal velocity.  We use the prescription
given in \citet{vink00}, which predicts wind terminal velocities on
the basis of the stellar mass, stellar radius, effective temperature,
and luminosity \citep[see also][]{lsl95}.  In turn, we interpolate the
stellar evolution calculations of Podsiadlowski for main sequence
stars to estimate effective temperatures.  Thus, we have not carried
out this calculation for stellar radii smaller than those on the
zero-age main sequence or larger than those on the terminal-age main
sequence (see Fig. \ref{fig:massfunc}).  We can compare these
``observed'' wind mass fluxes with those expected for a star with
given mass, effective temperature, and luminosity.  The ``expected''
mass flux is computed using formulae also presented by Vink et al.
These formulae are based on empirical fits of mass fluxes inferred
from ultraviolet, radio, and H$\alpha$ observations of many O and B
stars.  The same wind terminal velocity which we use to obtain an
estimate of the observed mass flux is used in the Vink et al.
prescription for wind mass flux.  The mass fluxes used by Vink et
al. in the empirical fits are for single stars, and are not, in
particular, for stars in binaries which nearly fill their Roche lobes.

We find that the ``observed'' wind mass loss rate is larger in all
cases than the ``expected'' mass loss rate.  In
Fig. \ref{fig:massfunc}, medium-size dots show those combinations of
system parameters in which the ``observed'' wind mass loss rate
exceeds the ``expected'' rate by a factor of 3 or less.  In the other
cases for which we computed the ``observed'' wind mass loss rate, it
exceeds the ``expected'' rate by more than a factor of 3, and often by
more than a factor of 10.  Most of these dots apply only to the $\beta
= 1$ case; one dot applies to the $\beta = 1/2$ case; none apply to
the $\beta = 0$ case.

The ``expected'' mass flux most closely approaches the ``observed''
mass flux for the case with $\beta = 1$, $M_c = 22.8 M_\sun$, $R_c =
20.0 R_\sun$, $v_{\infty} \sim 800$ km s$^{-1}$ for which we obtain
$\dot M_{\rm obs} \sim 2.1 \times 10^{-6} M_\sun$ yr$^{-1}$ and $\dot
M_{\rm exp} \sim 1.5 \times 10^{-6} M_\sun$ yr$^{-1}$.  Note that this
case is not realistic because the companion radius is equal to the
radius of its Roche lobe.  It is likely that the companion underfills
its Roche lobe; otherwise the mass accretion rate would be higher and
the X-ray luminosity would be expected to be at the Eddington limit.

\begin{deluxetable}{lccc}
\tablecaption{System Parameters}
\tablehead{
\colhead{Parameter}   &  \colhead{$\beta = 0$}  &  \colhead{$\beta = 1/2$} 
  & \colhead{$\beta = 1$} }
\startdata
$i$ & $54\arcdeg - 72\arcdeg$ & $43\arcdeg - 70\arcdeg$ & $38\arcdeg - 68\arcdeg$ \\
$a$ (lt-s) & $58 - 65$ & $58 - 75$ & $59 - 81$ \\
$M_c$ $(M_\sun)$ & $9 - 14$ & $9 - 22$ & $10 - 31$ \\
$R_c$ $(R_\sun)$ & 4 - 15 & 4 - 19 & 4 - 22 \\
$N_{H_{\rm wind}}(\phi=0)\tablenotemark{a}~~(10^{22}$ cm$^{-2}$) & 3.4 - 6.8 & 3.1 - 6.9 & 2.5 - 6.8 \\
$\dot{M}$ $(10^{-6} M_\sun$ yr$^{-1})$ & 2.3 - 15 & 1.9 - 15 & 1.3 - 14 \\
\enddata
\label{sysparam}
\tablenotetext{a}{The column density of the wind in the radial
direction outward to infinity from the position of the neutron star.}
\tablecomments{Each parameter range is defined by $\chi^2 < (1 +
  4.61/65)\chi^2_{min}$ for the fits of eq. 7 to the measured column
  densities.  The parameter ranges are also restricted by the
  requirements that $R_c \le R_{eclipse}$, $R_c \le R_{Roche}$, and by
  $R_{ZAMS} \le R_c \le R_{TAMS}$ for the associated value of $M_c$,
  i.e., the radius of the companion star is required to be in the
  range of radii of the main sequence stars calculated by
  Podsiadlowski (see Figs. \ref{fig:chisqr} and \ref{fig:massfunc}).}
\end{deluxetable}


\section{DISCUSSION}

We have found 605 s pulsations in the X-ray intensity of X1908+075.
Doppler shifts of the pulse frequency, and changes in the intensity and
low-energy attenuation that cyclically recur at the previously known
orbital period of 4.4 days allow us to measure orbital parameters and
to conclude that the system contains a highly magnetized neutron star
orbiting in the wind of a massive companion star.

The observed variations in the shape of the spectrum and in the
overall intensity indicate that the X radiation from the neutron star
is both absorbed and scattered in the wind.  We have estimated the
absorption on the basis of a crude spectral model that assumes that
any attenuation is caused by photoelectric absorption in neutral
(unionized) gas with solar element abundances.  In this model, we did
not include the effects of the photoionization of the wind by the
X-rays.  We can now use our estimates of the wind column density and
the X-ray luminosity to roughly estimate the ionization parameter $\xi
= L_X/n r^2$ \citep*{tts69} at a typical location in the binary
system, where $n$ is the number density of atoms in the wind and $r$
is the distance from the X-ray source.  If we take the orbital radius
$a$ as a characteristic distance within the binary system, i.e., we
set $r = a$, and for a density use $n \sim N_{H_{\rm wind}}(\phi =
0)/a$, we obtain $\xi = L_X/(N_H a) \sim$ 20--60 ergs cm s$^{-1}$.
Under these conditions we would expect that most atoms would lose
their outer electrons, but atoms of carbon, oxygen, etc. would retain
at least their K-shell electrons \citep{tts69,km82}.  \citet{kk84}
have estimated the equilibrium conditions of gases photoionized by
continuum X-ray spectra.  They consider the condition of the gas as a
function of a modified ionization parameter $\Xi = L_X/(4\pi c n r^2
kT)$ where $c$ is the speed of light, $k$ is Boltzmann's constant and
$T$ is the gas temperature.  For a characteristic location in the
X1908+075 system, we have $\Xi \sim 8 (T/10^5 {\rm K})^{-1}$.  For
each of the two spectral shapes that \citet{kk84} consider in detail,
one can roughly sketch the parameters of a self-consistent solution
taking into account that the actual density will be higher than that
estimated on the basis of the neutral matter absorption cross section.
The parameters of the solutions in the two cases are roughly the same:
$T \sim 4 \times 10^4$ K, $\Xi \sim 10$, the X-ray opacity in the
$\sim$2--6 keV band is reduced by a factor of $\sim 2$ from the value
expected for cold neutral gas of cosmic abundances, and $N_{H_{\rm
wind}}(\phi = 0) \sim 1 \times 10^{23}$ cm$^{-2}$.  Since, in this
regime, the temperature is a steep function of $\Xi$, it is likely to
vary over a wide range at different places in the system.  A more
realistic estimate of the physical properties of the wind would need
to self-consistently consider the actual source X-ray spectrum, the
wind velocity, the possibility of non-solar abundances, the
position-dependent degree of ionization in the wind, the effects of
the ionization upon the wind acceleration, and the transfer of
radiation in the system including both the effects of scattering and
absorption.  This is beyond the scope of this paper.

Using our fitted column densities and wind speeds estimated from the
formula of \citet{vink00}, we find the mass loss rate in the wind from
the companion star to be $\gtrsim 1.3 \times 10^{-6} M_\sun$ yr$^{-1}$
for all the cases with acceptable values of $\chi^2$ and with
companion star radii in the range expected for main sequence stars.
For those cases where the estimated mass loss rate is not more than a
factor of order 3 greater than an empirical prediction based on the
stellar mass, temperature, and luminosity, we find the rate must be
$\gtrsim 2 \times 10^{-6} M_\sun$ yr$^{-1}$.  The depression of the
X-ray opacity of the gas from ionization implies that the wind mass
loss rate must be higher than this latter rate, likely, as discussed
above, by a factor of two and possibly by a larger factor.

As noted in \S 5, over the allowed region of the $M_c - R_c$ plane
that we have identified, such mass loss rates are significantly higher
than those predicted using the empirically-based mass loss rate
prescriptions of Vink et al.  However, one should note that these
prescriptions do not take into account any of the effects of the star
being in a binary system, including the presence of a critical
potential lobe.  Furthermore, we used the effective temperatures
computed by Podsiadlowski for stars that evolve without mass loss and
without being affected by any of the phenomena that occur in a binary
system.  Thus, we cannot exclude the possibility that the companion
star is on the main sequence.


Another possibility is that the companion is a Wolf-Rayet (WR) star.
A WR star could well have a mass that is consistent with the value
that we find for the companion, but its radius would be much smaller
than that of a main-sequence star of comparable mass.  It would have a
prodigious wind \citep{nug00} that could reach or surpass the mass
loss rate that we infer (with assumptions) for X1908+075.  According
to \citet{nug00}, WR stars of either type WN or WC and mass in the
range $\sim 9$--15 M$_\sun$ have wind mass loss rates between 4 and
$20 \times 10^{-6}$ M$_\sun$ yr$^{-1}$.  Of course, the winds of WR
stars, which have little or no hydrogen, would have significantly
larger X-ray absorption cross sections per unit mass than winds from
unevolved stars.  If there is little hydrogen in the wind, we may have
{\em overestimated} the wind mass loss rate, but such a composition
would nonetheless clearly indicate the WR nature of the companion.

If the companion star is indeed a WR star, then the X1908+075 system
has potentially important implications, in general, for binary stellar
evolution, and for the formation, in particular, of neutron
star--black hole (NS-BH) systems.  If it is a WR star, then it is
likely the He or CO remnant core of a star that was originally much
more massive, possibly of mass $\sim 23-35~M_\sun$
\citep[e.g.,][]{hurl00}.  In this case we expect the companion to
undergo core collapse in $10^4$ to $10^5$ years and leave behind a
stellar-mass black hole \citep{brwn01}.  Thus, there is a possibility
that the X1908+075 system is the progenitor of a neutron star--black
hole binary where the neutron star formed first.  This would be
significant in at least two ways.  First, in spite of the fact that
six NS-NS binaries have been discovered \citep[see,
e.g.,][]{bur03,chmp04}, no NS-BH binaries have yet been found.  There
are theoretical arguments \citep{spn04,ppr04} which suggest that the
latter binaries should be at least a factor of 10 less populous than
their NS-NS cousins \citep[but see][]{vt03}. An identified progenitor
would help theorists better estimate the current population of NS-BH
binaries.  Second, if NS-BH binaries are formed, the issue of which
collapsed star forms first is also important to our understanding of
binary stellar evolution. The most direct way of producing these
systems is for the BH to form first \citep[from the more massive
component of the system; see,
e.g.,][]{pzve97,blwb00,fry01,nele01,lbw02,prh02}, but there are also
channels where the NS forms first, as discussed above \citep[see,
e.g.,][]{vt03}.

Since this paper was originally written, \citet{MG04} have reported
the results of near-infrared observations of stars in or close to the
error box of X1908+075, and have found a star whose JHK magnitudes and
colors, and H and K band spectra, suggest an O or B supergiant at $d
\sim 7$ kpc.  They believe that this star is likely to be the
counterpart of X1908+075.  If confirmed, this identification would
rule against a WR companion star.  We expect to obtain observations of
X1908+075 with the Chandra X-ray Observatory, and to thereby obtain an
X-ray position sufficiently accurate to secure the optical
identification.

\acknowledgements

We thank Edward Morgan for assistance with the data preparation.  We
are grateful to Ph. Podsiadlowski for providing us with mass-radius
relations for stars on the ZAMS and TAMS and for numerous helpful
discussions. We are also grateful to Henny Lamers for giving us useful
advice on stellar winds, and to an anonymous referee for helpful
comments.  One of us (SR) acknowledges support from NASA ATP Grant
NAG5-12522.
	


\begin{thebibliography}{}


\bibitem[Bildsten et al.(1997)]{bild97}
  Bildsten, L., Chakrabarty, D., Chiu, J., Finger, M. H., Koh, D., Nelson, R. W.,
  Prince, T. A., Rubin, B. C. et al. 1997, \apjs, 113, 367

\bibitem[Brown et al.(2001)]{brwn01}
  Brown, G.E., Heger, A., Langer, N., Lee, C.-H., Wellstein, S., \& 
  Bethe, H.A. 2001, New Astr., 6, 457

\bibitem[Brown et al.(2000)]{blwb00}
  Brown, G.~E., Lee, C.-H., Wijers, R.~A.~M.~J., \& Bethe, H.~A.\
  2000, \physrep, 333, 471

\bibitem[Burgay et al.(2003)]{bur03}
  Burgay, M., et al. 2003, \nat, 426, 531

\bibitem[Castor, Abbott, \& Klein(1975)]{castor75}
  Castor, J.~I., Abbott, D.~C., \& Klein, R.~I.\ 1975, \apj, 195, 157 

\bibitem[Champion et al.(2004)]{chmp04}
  Champion, D.~J., Lorimer, D.~R., McLaughlin, M.~A., Cordes, J.~M., Arzoumanian, Z., 
  Weisberg, J.~M., \& Taylor, J.~H. 2004, \mnras, 350, L61 [astro-ph/0403553]


\bibitem[Delgado-Mart{\'i} et al.(2001)]{del01}
  Delgado-Mart{\'i}, H., Levine, A. M., Pfahl, E., \& Rappaport, S. A.
  2001, \apj, 546, 455


\bibitem[Fryer \& Kalogera(2001)]{fry01}
  Fryer, C.L., \& Kalogera, V. 2001, \apj, 554, 548


\bibitem[Groenewegen \& Lamers(1989)]{gl89}
  Groenewegen, M.~A.~T.~\& Lamers, H.~J.~G.~L.~M.\ 1989, \aaps, 79, 359 


\bibitem[Hurley, Pols, \& Tout(2000)]{hurl00}
  Hurley, J.R., Pols, O.R., \& Tout, C.A. 2000, \mnras, 315, 543

\bibitem[{{Jahoda} {et~al.}(1996){Jahoda}, {Swank}, {Giles}, {Stark},
  {Strohmayer}, {Zhang}, \& {Morgan}}]{xte96}
  {Jahoda}, K., {Swank}, J.~H., {Giles}, A.~B., {Stark}, M.~J., {Strohmayer}, T.,
  {Zhang}, W., \& {Morgan}, E.~H. 1996, \procspie, 2808, 59


\bibitem[Joss \& Rappaport(1984)]{jr84}
  Joss, P.~C.~\& Rappaport, S.~A.\ 1984, \araa, 22, 537 


\bibitem[Kallman \& McCray(1982)]{km82}
  Kallman, T.R., \& McCray, R. 1982, \apjs, 50, 263

\bibitem[Krolik \& Kallman(1984)]{kk84}
  Krolik, J.H., \& Kallman, T.R. 1984, \apj, 286, 366



\bibitem[Lamers \& Cassinelli(1999)]{iswbook99}
  Lamers, H.~J.~G.~L.~M.~\& Cassinelli, J.~P.\ 1999, Introduction to stellar winds / 
  Henny J.G.L.M.~Lamers and Joseph P.~Cassinelli.~Cambridge ; New York : 
  Cambridge University Press, 1999.~ ISBN 0521593980

\bibitem[Lamers, Snow, \& Lindholm(1995)]{lsl95}
  Lamers, H.J.G.L.M., Snow, T.P., \& Lindholm, D.M. 1995, \apj, 455, 269 

\bibitem[Lee, Brown, \& Wijers(2002)]{lbw02}
 Lee, C., Brown, G., \& Wijers, R. 2002, \apj, 575, 996

\bibitem[Lucy \& Solomon(1970)]{lucy70}
  Lucy, L.~B.~\& Solomon, P.~M.\ 1970, \apj, 159, 879 

\bibitem[Morel \& Grosdidier(2004)]{MG04}
  Morel, T. \& Grosdidier, Y. 2004, \mnras, submitted 

\bibitem[Nagase(1989)]{nag89} 
  Nagase, F.\ 1989, \pasj, 41, 1 

\bibitem[Nelemans \& van den Heuvel(2001)]{nele01}
  Nelemans, G., \& van den Heuvel, E.P.J. 2001, \aap, 376, 950


\bibitem[Nugis \& Lamers(2000)]{nug00}
  Nugis, T., \& Lamers, H.J.G.L.M. 2000, \aap, 360, 227

\bibitem[Pfahl, Podsiadlowski, \& Rappaport(2004)]{ppr04}
  Pfahl, E., Podsiadlowski, Ph., \& Rappaport, S. 2004, in preparation



\bibitem[Podsiadlowski, Rappaport, \& Han(2002)]{prh02}
  Podsiadlowski, Ph., Rappaport, S., \& Han, Z. 2002, \mnras, 341, 385


\bibitem[Sipior, Portegies Zwart \& Nelemans(2004)]{spn04}
  Sipior, M. S., Portegies Zwart, S., \& Nelemans, G., astro-ph/0407268

\bibitem[Portegies Zwart, Verbunt, \& Ergma(1997)]{pzve97}
  Portegies Zwart, S.F., Verbunt, F., \& Ergma, E. 1997, \aap, 321, 207

\bibitem[Puls et al.(1996)]{puls96}
  Puls, J., et al.\ 1996, \aap, 305, 171 


\bibitem[Rothschild et al.(1998)]{Roth98}
  Rothschild, R. E., et al. 1998, \apj, 496, 538

\bibitem[Shull \& van Steenberg(1985)]{ss85} 
  Shull, J.~M.~\& van Steenberg, M.~E.\ 1985, \apj, 294, 599 

\bibitem[Sobczak et al.(2000)]{sob00}
  Sobczak, G. J., McClintock, J. E., Remillard, R. A., Cui, W.,
  Levine, A.  M., Morgan, E. H., Orosz, J. A., \& Bailyn, C. D. 2000,
  \apj, 544, 993

\bibitem[Tarter et al.(1969)Tarter, Tucker, \& Salpeter]{tts69}
  Tarter, C.B., Tucker, W.H., \& Salpeter, E. E. 1969, \apj, 156, 943


\bibitem[Vink, de Koter, \& Lamers(2000)Vink et al.]{vink00}
  Vink, J.S., de Koter, A., \& Lamers, H.J.G.L.M. 2000, \aap, 362, 295

\bibitem[Voss \& Tauris(2003)]{vt03}
 Voss, R., \& Tauris, T.M. 2003, \mnras, 342, 1169


\bibitem[Wen, Remillard, \& Bradt(2000)]{WRB}
  Wen, L., Remillard, R.A., \& Bradt, H.V. 2001, \apj, 532, 1119

\end{thebibliography}
\end{document}